\documentclass[aps,prb,superscriptaddress,twocolumn]{revtex4-1}

\usepackage{amsmath, amsthm, amssymb}
\usepackage{graphicx}
\usepackage{color}
\usepackage{times}

\usepackage{natbib}
\usepackage{hyperref}
\hypersetup{
        colorlinks=true,
}

\newtheorem{theorem}{Theorem}
\newtheorem{conjecture}{Conjecture}
\theoremstyle{definition}
\newtheorem{definition}[theorem]{Definition}

\newcommand{\eqnref}[1]{(\ref{#1})}
\newcommand{\ord}[1]{\mathcal{O}(#1)}
\newcommand{\Hint}{H_{\text{int}}}
\newcommand{\baa}{Basko {\it et al.} \cite{Basko06a,Basko06b}\ }
\newcommand{\baaa}{Basko {\it et al.} \cite{Basko06a,Basko06b}}
\newenvironment{mypar}[2]
  {\begin{list}{}%
    {\setlength\leftmargin{#1}
    \setlength\rightmargin{#2}}
    \item[]}
  {\end{list}}

\newcommand{\Tr}{\text{Tr}}

\begin{document}

\title{Area laws in a many-body localized state and its
implications for topological order}

\author{Bela Bauer}
\affiliation{Station Q, Microsoft Research, Santa Barbara, CA 93106-6105, USA}

\author{Chetan Nayak}
\affiliation{Station Q, Microsoft Research, Santa Barbara, CA 93106-6105, USA}
\affiliation{Physics Department, University of California,  Santa Barbara, CA 93106, USA}

\begin{abstract}
The question whether Anderson insulators can persist to finite-strength
interactions -- a scenario dubbed many-body localization -- has recently
received a great deal of interest. The origin of such a many-body localized phase
has been described as localization in Fock space, a picture we examine numerically.
We then formulate a precise sense in which a {\it single energy eigenstate}
of a Hamiltonian can be adiabatically connected to a
state of a non-interacting Anderson insulator. We call such a state
a {\it many-body localized state} and define a many-body localized
{\it phase} as one in which almost all states
are many-body localized states.
We explore the possible consequences of this;
the most striking is an area law for the entanglement entropy of
almost all excited states in a many-body localized phase.
We present the results of numerical calculations for a one-dimensional
system of spinless fermions. Our results are consistent with an
area law and, by implication, many-body localization for almost
all states and almost all regions for
weak enough interactions and strong disorder. However, there are
rare regions and rare states with much larger entanglement entropies.
Furthermore, we study the implications that many-body localization
may have for topological phases and self-correcting
quantum memories. We find that there are scenarios in which many-body localization can
help to stabilize topological order at non-zero energy density,
and we propose potentially useful criteria to confirm these scenarios.
\end{abstract}

\maketitle

\section{Introduction}

Until relatively recently, it was generally assumed that all materials
have non-vanishing electrical conductivity at non-zero temperature
even in the presence of disorder.
According to this conventional wisdom, an electron trapped in some
potential energy well would always have some non-zero probability
of being thermally-excited to a nearby site and, from there, to another,
and so on. However, \baa suggested, following ideas of Anderson \cite{Anderson58}, that it could, instead, be possible for a system
to remain an insulator even at non-zero temperature,
an effect that they called {\it many-body localization} (MBL).
According to their analysis, a weakly-interacting system of
localized electrons cannot serve as its own heat bath (although phonons,
which would necessarily be present in a solid but not in a system
of ultra-cold atoms, could serve in that role).
Consequently, the system cannot supply an electron with the energy
that it needs to make a transition to a nearby site even when
there is non-zero {\it energy density}, so that the system
has far more total energy than is needed for this transition.

This hypothesis has profound consequences, not only for the electrical
conduction of dirty metals and semiconductors (where, due to phonons,
there will always be non-zero conduction at non-zero temperature,
though it would be very small
in a many-body localized state in the limit of weak electron-phonon interaction
\cite{Basko07}), but also for the foundations of quantum statistical mechanics.
The eigenstate thermalization hypothesis (ETH) \cite{Deutsch91,Srednicki94} is an elegant
justification of the microcanonical ensemble: expectation values of physical
observables in individual energy eigenstates in generic (i.e. non-integrable) systems
are equal to their averages over all eigenstates of that energy -- i.e. are equal
to the predictions of the microcanonical ensemble. If, in a closed isolated system
in an energy eigenstate, one focuses on any subsystem
which is large but still much smaller than the whole system,
the rest of the system can act as a heat bath, and the system
will have the properties expected in thermal equilibrium.
However, this must fail in systems in which many-body localization occurs.
The rest of the system cannot act as a heat bath; rather than thermalize
any added energy, it keeps it localized.

\baa proposed many-body localization
as an analogue of single-particle localization, in which a single
particle moving on a lattice with (for instance) random on-site energies is unable,
in an energy eigenstate, to explore the whole lattice and, instead,
has an amplitude that decays exponentially with distance from
some finite subset of the lattice. Similarly, if a many-body state
has low energy density, which corresponds to low temperature,
then it is unable to explore the entire Fock space of Slater
determinants of single-particle eigenstates of the non-interacting
disordered system and, instead, has appreciable amplitude
only on some subspace.
This picture of localization in Fock space has previously been discussed
for quantum dots\cite{Altshuler1997,Berkovits1998,Monasterio1998}.

Important progress was made by Oganesyan and Huse \cite{Oganesyan07},
who noted that in a system with a maximum energy, such as a system
of fermions on a lattice with a finite number of bands, it could be possible
for many-body localization to occur even at ``infinite temperature'', meaning
that it would hold for all energy eigenstates. They studied the energy
level statistics of a system of interacting fermions in one dimension,
and found that the energies, obtained by numerical diagonalization
of the Hamiltonian, obeyed Poisson statistics for weak interactions
and random-matrix level statistics
for stronger interactions. The former corresponds to a many-body localized
phase and the latter to a phase with extended states. However, the
transition point between these two regimes drifted with system size,
leaving open the possibility that all states are extended in the
thermodynamic limit. The ``infinite temperature'' ensemble was further
explored in Ref.~\onlinecite{Pal2010}.

Other probes of many-body localization, primarily involving
the real-time dynamics of systems prepared in simple initial states,
were investigated in a number of papers.
Iyer {\it et al.} \cite{Iyer12} computed the density auto-correlation function,
configuration space participation ratio, and growth of the Renyi entanglement
entropy in clean but quasiperiodic systems prepared in initial product states
in the local occupation basis. They found two regimes, one in which
these measures indicated that the subsequent evolution
explored the entire Hilbert space, and another, interpreted as a
many-body localized phase, in which it does not.
\u{Z}nidari\u{c} et al.~\cite{Znidaric08},
De Chiara et al.\cite{DeChiara2006},
Bardarson et al. \cite{Bardason12}
and Vosk and Altman \cite{Vosk13} studied the growth of the
entanglement entropy in the evolution of the system from an initial
product state in the local occupation basis. They
found logarithmic growth, which was also interpreted as a possible indication
of many-body localization. Theoretical explanations for this growth were
recently discussed in Refs.~\onlinecite{serbyn2013-1,Huse2013}. A logarithmic bound was also shown
in Ref.~\onlinecite{Burrell2007} for the non-interacting case.
(See Sec.~\ref{sct:discussion} for further discussion of the
growth of the entropy from an initial product state.)

In this paper, we put the \baa
picture to the test by numerical simulation
of a system of spinless electrons with weak nearest-neighbor
interactions in one dimension.
In Section~\ref{sct:fock}, we show that, if we view the Slater determinants of (localized)
single-particle wavefunctions as the nodes of a graph, then
the matrix elements of the Hamiltonian only connect states that
differ by local rearrangements of one or two electrons and
are nearby in energy. Thus, motion on this graph occurs
by ``local'' hops, roughly analogous to the motion of a single
particle on a lattice. However, the coordination number
of this graph scales linearly with the system size;
in other words, lowest-order perturbation theory about
the non-interacting state gives a contribution that is $O(N)$,
where $N$ is the number of particles. In this respect, the many-body
system seems different from the single-particle one, in which
localization occurs in the strong disorder limit because
perturbation theory in the hopping converges \cite{Anderson58,Frohlich83}.
However, the failure of this extremely literal analogy between
many-body localization and single-particle localization can be
rather benign. Since, in a typical energy eigenstate, each particle
can hop to a nearby state, there are $O(N)$ states
with which a Slater determinant will have appreciable matrix element.
Therefore, an energy eigenstate of an interacting system will
generically not be a superposition
of a finite number of eigenstates of the non-interacting system.
However, the Hilbert spaces of many-body systems
have dimension of $\ord{e^{bN}}$, which is so
large that, even if a state explores $\ord{e^{aN}}$ states with $a<b$,
this is still a very small fraction of the full Hilbert space.

In fact, we will argue in this paper that a system in a many-body
localized phase explores only a relatively small and very specific subset of the
full Hilbert space, namely the low-entanglement states.
We will define many-body localization for an
eigenstate of such a system as a state that may, in the following sense,
be adiabatically connected to a state of the non-interacting system, which is
a Slater determinant of localized single-particle states:
\begin{mypar}{0.5cm}{0.5cm}
For a many-body localized energy eigenstate $|\psi\rangle$,
there is a finite-depth local unitary transformation $U$
that will almost everywhere transform $|\psi\rangle$ into
a Slater determinant of localized single-particle states,
to within desired accuracy.
\end{mypar}
In other words, in a generic region of the lattice,
the state looks similar to a product state.
In Section \ref{sct:entropy}, we
give a more precise version of this definition.
Note that this is a definition that
applies to {\it a single eigenstate}. The surprise is
that there are excited states with a finite energy density
above the ground state that satisfy this definition. We expect that
the {\it ground state} of a disordered system in one or two dimensions
(or, with sufficient disorder, a higher-dimensional system)
will satisfy the definition given above
(denoted Definition \ref{def:MBL-state}
in Section \ref{sct:mbl-def}) even for strong interactions.
However, for very weak interactions and strong disorder,
the system may be in a regime in which almost all energy eigenstates
satisfy this definition. We believe that this is closely related to the concept of
``many-body localization at infinite temperature'',
discussed by Oganesyan and Huse \cite{Oganesyan07}.
We will say that such a system is in a {\it many-body localized phase}.
It has been speculated that, for intermediate interactions, there is
a critical energy density below which all states are many-body localized.
However, we do not find evidence for a sharp energy density that is analogous to the mobility edge of single-particle localization in
spatial dimensions $d>2$.

A possible consequence of Definition \ref{def:MBL-state},
which we discuss in more detail
in Section~\ref{sct:area-law}, is that an energy eigenstate
satisfying it will have a bipartite entanglement entropy with an `area law':
for almost all regions of size $L$ larger than the correlation
length, the entanglement entropy $S$ between the region and the
rest of the system has leading $L$-dependence
\begin{equation}
\label{eqn:area-law}
S = \alpha L^{d-1} + \ord{L^{d-2}},
\end{equation}
where
$\alpha$ is a constant, independent of $L$, and $d$ is the spatial dimension.
This is in sharp contrast to random quantum states, or highly excited (thermal)
states of generic local Hamiltonians, which will obey a volume-law scaling\cite{Eisert2010}.
For the ground state of a disordered, interacting system, such an area law was previously
observed in Ref.~\onlinecite{Berkovits12}. Here, however, we emphasize that
in a many-body localized phase, an area law holds for nearly all eigenstates
instead of merely for low-energy states.
A further consequence is that
the mutual information, which serves as an upper bound for
all correlation functions\cite{Wolf2008}, decays exponentially.
Note that entanglement properties have previously been used to
identify phase transitions~\cite{Vidal2003}, including transitions
in disordered systems\cite{Chakravarty10}.

In our definition of an MBL state and our statement of the
area law, we used the terms ``almost everywhere'' and ``almost
all regions'' because we find that there are rare regions in the system
where the entanglement entropy is large. Furthermore,
we wrote ``nearly all eigenstates'' in the previous paragraph because
we find that there are states of arbitrarily high entanglement entropies
at all energies, but the density of such states is exponentially-small.
This is one reason why it is difficult to identify a sharp
`mobility edge' separating high and low
entanglement entropy states.

We find clear evidence for two regimes in
a system of interacting spinless fermions in 1D.
In the weak-interaction regime, the median value of
$S$ does not scale with $L$
(as expected from Eq. (\ref{eqn:area-law}) with $d=1$).
The density of states with high entanglement entropy falls off
exponentially. Thus, our results are consistent with the existence of
many-body localized states in this regime.
In the strong-interaction regime, the median value of $S$ increases with
$L$ in a manner consistent with a linear dependence on $L$
and the distribution of entanglement entropies is a Gaussian peaked
near the median value. The situation is less clear
for intermediate interaction strengths. It is possible that
there is a sharp transition between the two regimes described above.
An alternative possibility is that there is an intermediate phase in which
there are comparable densities of high and low entropy states.
We cannot, at present, rule out either of these possibilities.

One interesting possible consequence of many-body localization is that
it may stabilize topological order in a manner analogous to the stabilization
of zero-temperature topological qubits by single-particle localization.
Here, we give a criterion for
the stabilization of 2D topological order by many-body localization
of quasiparticles. Our criterion relies on the use of Wilson loops. However,
a related criterion in terms of entanglement entropy can also be formulated.
This is potentially particularly useful when the topological phase
is dual to the symmetry-unbroken phase of a magnetic system.
Similar ideas were recently discussed in a paper by
Huse {\it et al.} \cite{Huse13}.

The rest of this paper is structured as follows:
In Section \ref{sct:fock}, we discuss a 1D model of interacting
spinless fermions in a random on-site potential from the perspective
of localization in Fock space. In Section \ref{sct:entropy}, we introduce
our definition of MBL in terms of continuation to a local product state
with a finite-depth unitary transformation. We show how this implies
an area law for the entanglement entropy and present calculations
of the entanglement entropy for our 1D model. In Section \ref{sct:topological}
we comment on some of the implications of these results
for topological order and quantum memories at non-zero energy density.
Finally, in Section \ref{sct:discussion}, we discuss our results.

\section{Localization in Fock Space}
\label{sct:fock}

\baa proposed that many-body
localization could occur as an analogue of single-particle localization,
in the sense that the state of the many-body system could be localized
in Fock space. According to their analogy,
the following correspondence holds:
\begin{eqnarray*}
\mbox{Lattice Site } &\leftrightarrow& \mbox{Slater
determinant  } \Psi^{\rm Sl}_\alpha\cr
& & \mbox{of localized single-particle}\cr
& & \mbox{energy eigenstates}\cr
\mbox{Random On-Site Potential} &\leftrightarrow&
\mbox{Hartree-Fock}\cr
& & \mbox{energy } \langle \Psi^{\rm Sl}_\alpha| H | \Psi^{\rm Sl}_\alpha\rangle\cr
& & \mbox{of Slater determinant  } \Psi^{\rm Sl}_\alpha\cr
\mbox{Hopping} &\leftrightarrow& \mbox{Interaction} 
\end{eqnarray*}
In this section, we briefly discuss this analogy, while we postpone a discussion of our numerical results to App.~\ref{sct:fock-app}. For another recent numerical study of this analogy, see Ref.~\onlinecite{Monthus2010}.

For the sake of concreteness, let us consider the following Hamiltonian in one dimension:
\begin{subequations} \label{eqn:H} \begin{align}
H &= H_0 + \Hint \\
H_0 &= -t \sum_{i=1}^{L-1} \left( c_i^\dagger c_{i+1} + c_{i+1}^\dagger c_i \right) + \sum_{i=1}^L w_i n_i \\
\Hint &= V \sum_{i=1}^{L-1} n_i n_{i+1},
\end{align} \end{subequations}
where $c^\dagger_i$ creates a spinless fermion on site $i$, and $n_i=c^\dagger_i c_i$. The $w_i$ are uniformly chosen from $w_i \in [-W,W]$. We use open boundary conditions. The non-interacting problem $H_0$ is readily solved to obtain the eigenvalues $\varepsilon_n$ and eigenvectors $\phi_n(i)$. We can then rewrite the problem as follows:
\begin{equation} \label{eqn:Hsp}
H = \sum \varepsilon_n d_n^\dagger d_n + \sum_{ijkl} V_{ijkl} d_i^\dagger d_j^\dagger d_k d_l,
\end{equation}
where $d_n^\dagger$ creates a fermion in the single-particle state $\phi_n(i)$. Note that in this formalism, the single-particle states $\phi_n(i)$ are obtained as solutions purely of the non-interacting problem and not as solutions of a Hartree-Fock equation for the interacting Hamiltonian. Finally, this can be recast in the following form:
\begin{equation} \label{eqn:Hfock}
H = \sum_{\vec{\alpha}} \mu_{\alpha} |\vec{\alpha}\rangle \langle \vec{\alpha} | +\sum_{\vec{\alpha}\neq\vec{\beta}} V_{\beta \alpha} |\vec{\beta}\rangle\langle\vec{\alpha}|.
\end{equation}
In this model, the on-site potential $\mu_\alpha$ is given by the Hartree-Fock energy of the single-particle Slater determinants, which consists of the non-interacting part of the original Hamiltonian as well as the diagonal part of the interaction term. The hopping term $V_{\beta \alpha}$ stems purely from the interaction term. Full definitions for all these models are given in App.~\ref{sct:fock-app}.

It was argued by \baa that Hamiltonian~\eqnref{eqn:Hsp} has a very particular structure, namely that due to the localization of the single-particle orbitals $\phi_n(i)$, the matrix elements $V_{ijkl}$ fall off exponentially with separation between the states $\phi_i$ and $\phi_k$
or $\phi_l$ and between $\phi_j$ and $\phi_k$ or $\phi_l$.
They further argued that these matrix elements decay very quickly with the difference between
single-particle energies.
This holds for a quantum dot in the diffusive regime, where quantum interference
corrections can be neglected \cite{Aleiner02,Mirlin00}, whereas we will be dealing
with the strong disorder regime in this paper. We therefore confirm this condition
numerically in App.~\ref{sct:fock-app}.

In order to be able to apply a single-particle localization perspective to the Fock space hopping problem~\eqnref{eqn:Hfock},
it is necessary to understand the properties of the (random) graph on which this hopping Hamiltonian is defined. To make
progress on this, we calculate an effective coordination number
\begin{align} \label{eqn:z}
z&=\left \langle z_\alpha \right\rangle \, ,&z_\alpha &= \sum_{\beta\neq\alpha} \frac{V_{\beta \alpha}}{|\mu_\beta - \mu_\alpha|}
\end{align}
where $\langle \cdot \rangle$ indicates averaging over $\alpha$ as well as disorder realizations. As discussed in more detail
in App.~\ref{sct:fock-app}, we observe that the effective coordination number scales linearly in the system size, $z \sim L$,
independently of the parameters of our model, i.e. in both the regime where we expect a delocalized phase and the
regime in which we expect an MBL phase. This divergence is easily understood by the following argument: in a generic
Slater determinant $|\vec{\alpha}\rangle$ at half-filling, there are $N=L/2$ fermions that can hop within a small part of
the system, whose volume is controlled by the localization length and essentially is $\xi_{\text{loc}}^d$. In our Fock space
model, the number of Slater determinants $|\vec{\beta}\rangle$ to which the system can ``hop" is therefore also proportional
to the number of fermions $N=L/2$, and hence our coordination number $z \sim N$. In App.~\ref{sct:fock-app}, we confirm
this argument by calculating the dependence of $z$ on the number of fermions, $N$, in a system of fixed size $L$.

We conclude that the effective coordination number scales as $N$.
This is not very deep: each of the $N$ electrons can make a transition to
a nearby (in both location and energy) state. However, this means that
MBL is not likely to be a simple analogue of single-particle localization,
with Eq.~(\ref{eqn:Hfock}) playing the role of $H_0$ in Eq.~(\ref{eqn:H})
for a single particle. This is further illustrated by consideration of the inverse
participation ratios:
\begin{equation}
I_\psi = \sum_{\alpha} |\langle \alpha | \psi\rangle|^4.
\end{equation}
Precisely at $V=0$, the inverse participation ratio is $1$ in every eigenstate.
As $V$ is increased, we find that the inverse participation ratios
decrease linearly with $V$.
But our calculated values of ${I_\psi}$ as a function of $V$
do not show simple scaling behavior as a function of system size.
Moreover, we expect that ${I_\psi} \sim e^{-aN}$ even in an MBL
phase because each electron can hop to $O(1)$ nearby localized states,
so it would be difficult to use ${I_\psi}$ to distinguish an MBL phase
from a metallic state.

\section{Adiabatic Continuity and Entanglement Entropy}
\label{sct:entropy}

\subsection{Definition of a Many-Body Localized State}
\label{sct:mbl-def}

In this paper, we adopt the point of view that
when all of the states of a system are many-body localized,
they should be adiabatically connected to the localized states
of the corresponding non-interacting system.
However, it is difficult to make this notion precise by considering
direct analogues of criteria used in single-particle localization,
such as inverse participation ratios,
since even weak interactions will mix
exponentially many single-particle Slater
determinants to obtain the eigenstate of the interacting system.

For this reason, the {\it definition} of many-body localization
that we propose in this paper is, as stated in the introduction:
\begin{mypar}{0.5cm}{0.5cm}
For a many-body localized energy eigenstate $|\psi\rangle$,
there is a finite-depth local unitary transformation $U$
that will almost everywhere transform $|\psi\rangle$ into
a Slater determinant of localized single-particle states,
to within desired accuracy.
\end{mypar}

In this section, we will try to make the words ``almost'' and
``to within desired accuracy'' more precise. For the most part,
we will assume (without justification) in this paper
that the latter caveat does not matter,
but the first one will play a role in our calculations. The first step
in making our definition more precise is to define the
``localization depth'' of a state
(for a definition of a local unitary transformation of depth $D$ and localized single-particle states, see Appendix~\ref{sec:technical-details}):
\begin{definition}
\label{def:loc-depth}
A state $|\psi\rangle$ on a lattice $\mathcal{L}$ has {\it localization depth} $D$ {\it to accuracy} $(\epsilon,k)$
in some (not necessarily connected) subset $\mathcal{A}$
if there exists a local unitary transformation $U$ of depth
$D$ and a Slater determinant of localized single-particle states $|\Psi^{\text{Sl}}\rangle$,
such that the reduced density matrices,
\begin{align} \nonumber
\rho_{\mathcal{B}} &= \Tr_{\mathcal{L} \setminus \mathcal{B}} |\psi\rangle\langle\psi|,
&\rho_{\mathcal{B}}^{\text{Sl}} &= \Tr_{\mathcal{L} \setminus \mathcal{B}} |\Psi^{\text{Sl}}\rangle\langle\Psi^{\text{Sl}}|,
\end{align}
satisfy the property that
\begin{equation} \nonumber
\Tr\left(\left|\rho_{\mathcal{B}} - U \rho_{\mathcal{B}}^{\text{Sl}} U^\dagger\right|\right) < \epsilon.
\end{equation}
for all connected $\mathcal{B}\subset\mathcal{A}$ with
$\text{vol}(\mathcal{B})=k$.
\end{definition}

Here, $\Tr(| \cdot |)$ denotes the trace norm distance, which for reduced density
matrices of two states has the operational meaning that it quantifies how well the two
states can be distinguished by measurements only on these reduced density matrices~\cite{Helstrom1976,Fuchs1999,Bauer2010}.
This definition thus formalizes the notion that such a state shares all $k$-local properties of a localized Slater determinant, i.e.
properties that can be measured using $k$-local operators.
A key requirement is that a {\it single} transformation $U$ and a {\it single} Slater determinant
$|\Psi^\text{Sl}\rangle$ are used for {\it all} $\mathcal{B}$; if one allows a different
$U$ or $|\Psi^\text{Sl}\rangle$ for each $\mathcal{B}$, the above
definition can always be fulfilled with $D \sim k$.
In our definition, we require that $|\Psi^\text{Sl}\rangle$ be a Slater determinant
of localized single-particle states, as defined in Appendix~\ref{sec:technical-details},
rather than the basis of site occupation numbers so that our unitary $U$ does not
have to remove the exponential tails of localized states, as we discuss further
in the next subsection.

Of course the localization depth $D$ will generally grow with $k$, i.e. as
more long-ranged properties of the system are being captured. A similar picture
emerges when describing gapped states in one dimension as local unitary circuits
applied to a product state, which is equivalent to describing them with a matrix-product
state of bond dimension $M \sim \exp(D)$: while local properties may be accurately described
by a circuit of finite depth $D$ even as $L \rightarrow \infty$~\cite{Verstraete2006}, the bond
dimension $M$ will have to grow polynomially with $N$ if some fixed accuracy is demanded
for the density matrix on $N$ sites~\cite{Hastings2007-prb,Hastings2007}. In our case,
the localization depth $D$ would depend on $\text{vol}(\mathcal{A})$ if one were to demand
$\mathcal{B} = \mathcal{A}$; in order to avoid this dependence and to facilitate approaching
the thermodynamic limit below, we only require $k$-local properties
to match to a localized Slater determinant, allowing us to obtain a finite $D$ independent of
$\text{vol}(\mathcal{A})$, for a given $k$.


Of course, the above definition can always be fulfilled if $D$ is allowed
to scale sufficiently rapidly with $\text{vol}(\mathcal{A})$ -- in other words,
in a finite-size system, every state has a finite localization
depth in all subsets. The distinction between many-body localized states
and extended ones must, therefore, be the scaling of the localization depth
with the system size. We now propose a definition of a many-body localized
state that draws a line between the two types of states.
For simplicity, we give a definition on the hypercubic lattice $\mathbb{Z}^d$,
but this definition can be generalized to an arbitrary lattice or even an
arbitrary triangulation.
\begin{definition}
\label{def:MBL-state}
We will say that an energy eigenstate $|\psi\rangle$
of a Hamiltonian $H$ on $\mathbb{Z}^d$
is an {\it MBL state} if,
for any $\epsilon>0$ and any
$0<f<1$, there exists a sequence of hypercubic regions
$C_i \subset \mathbb{Z}^d$
satisfying ${C_1} \subsetneq {C_2} \subsetneq \ldots$
and a sequence of subsets ${\mathcal A}_i \subset C_i$
satisfying ${\mathcal A}_1 \subsetneq {\mathcal A}_2 \subsetneq \ldots$ and $\text{vol}({{\mathcal A}_i})/ \text{vol}({C_i}) > f$ such that the following holds:
$D\equiv\lim_{i\rightarrow\infty} D_{i}$
is finite for any finite $k$,
where $D_{i}$ is the localization depth of
$|\psi\rangle$ to accuracy $(\epsilon,k)$ on ${\mathcal A}_i$.
\end{definition}

In other words, an MBL state can be approximately transformed into a Slater determinant of localized single-particle orbitals almost everywhere by a
local unitary transformation of finite
depth. We emphasize that, at this point, our definition applies to a single eigenstate of a Hamiltonian.
The details of the Hamiltonian determine whether any of its eigenstates are MBL states. However, these details -- e.g. whether there is a random potential, whether the interactions are short-ranged or long-ranged, etc. -- do not enter our definition of an MBL state.
To the best of our knowledge, such a definition has not been previously formulated. (However, see Refs.~\onlinecite{Huse2013,Serbyn2013}
where similar ideas are discussed.) Most previous discussions have focused on the dynamical properties of Hamiltonians. Our treatment
can be more directly compared to those discussions
when, in Section \ref{sec:MBL-phase-def}, we
give a definition of a many-body localized phase that applies directly
to the properties of a Hamiltonian.

This definition provides an attempt to formalize the basic idea
that states of a many-body localized system are adiabatically
connected to states of a non-interacting Anderson insulator.
A non-interacting Anderson insulator
has the unusual property that {\it all} of its states below the mobility edge
are Slater determinants of {\it localized} single-particle orbitals.
While this is also true for the ground
state of a band insulator (since fully occupied
bands can be expanded in localized Wannier orbitals),
it is not true for its excited states at non-zero energy density.
It is also not true for arbitrary eigenstates (including the ground state) of
symmetry-protected topological phases~\cite{gu2009}$^,$\footnote{Note that for an SPT phase, such a decomposition is possible if one first applies a finite-depth unitary transformation that breaks the symmetry.}
or gapless phases.
Our definition essentially boils down, then, to the idea that
this unusual property of a non-interacting Anderson insulator
can also hold, to within desired accuracy, in an interacting system.
Note that our definition is not precisely the same as a failure
of the ETH hypothesis \cite{Deutsch91,Srednicki94};
one can imagine non-ergodic states that aren't MBL states \cite{Rigol12}.

In this respect, we can connect Definition~\ref{def:MBL-state} to a definition given by \baa in
their pioneering work. Although our definition applies to an individual energy eigenstate
rather than a range of temperatures and focusses on
the relation to a Slater determinant via local unitary transformations
of finite-depth, it has similar consequences if it holds for almost all states
of a system.
Consider the following
statement of many-body localization given by \baaa:
Label the energy eigenstates of the system $|k\rangle$ and consider
matrix elements of a bounded local operator $X$,
\begin{equation}
C_{k' k} \equiv \langle k'| X | k \rangle
\end{equation}
(\baa consider the operator $c^\dagger_\alpha c_\beta$,
but their statement can be generalized to any local operator $X$.) Then, one could
define MBL by requiring that if all $|k\rangle, |k'\rangle$ are MBL states
in subset $\mathcal{A}$, then
\begin{equation}
\label{eqn:matrix-elts-local-ops}
\frac{\sum_{k'} |C_{k' k}|^4}{(\sum_{k'} |C_{k' k}|^2)^2}
\end{equation}
has a non-zero limit in thermodynamic limit, while in the metallic phase,
it has a vanishing limit. The sum in Eq.~\ref{eqn:matrix-elts-local-ops}
will have a non-zero limit if there is one term in the sum
that does not scale with system size. This property can be inherited
from the non-interacting case as follows.
We can insert the local unitary
transformation $U_k$ for each state $|k\rangle$
to obtain
\begin{equation}
C_{k' k} = \langle k'| U_{k'}^\dagger U_{k'} X U_k^\dagger U_k |k\rangle =  \langle\Psi^{\rm Sl}_{k'}| U_{k'} X U_k^\dagger |\Psi^{\rm Sl}_{k}\rangle
\end{equation}
If $X$ has support only on $\mathcal{A}$ and $\mathcal{A}'$,
the regions of the two states where $U_k$, $U_{k'}$ have finite-depth,
then $U_{k'} X U_k^\dagger$ is also
a local operator and $|\Psi^{Sl}_{k}\rangle$ is a Slater
determinant of localized single-particle wavefunctions.

\subsection{Area-Law for the Entanglement Entropy}
\label{sct:area-law}

A key proposition of this paper is that many-body localized states at non-zero energy,
which meet the requirements of Definition~\ref{def:MBL-state}, have an area law
for the entanglement entropy.
This area law will lead to a more practically useful definition for an MBL state
since it is generically difficult to find the finite-depth local unitary
transformation $U$ required by Definition~\ref{def:MBL-state}, whereas
entanglement properties are readily calculated by a variety of
computational and analytical approaches.

In Section \ref{sct:fock}, we have encountered the issue that even a localized
system may explore an exponentially large part of its Hilbert space.
The same issue is also encountered when attempting to
accurately describe the low-energy states of
a gapped system such as a band insulator.
Expanding a low-energy state of such a system in
a local basis, one finds contributions from exponentially-many
local product states. At the same time,
it is known that the ground states of band insulators
-- in fact, the ground states of most local Hamiltonians --
only have overlap with a tiny subset of the full Hilbert space,
namely that of low-entanglement states,
which obey an area law~\cite{Bekenstein1973,Eisert2010,Bombelli1986,Srednicki1993}
or, in the case of many gapless systems,
an area law with a multiplicative logarithmic correction.
In one dimension, the area law
for the ground state of systems with a finite correlation length
has been firmly established~\cite{Hastings2007},
while a logarithmic correction is expected for the ground state of
critical (scale-invariant) systems~\cite{holzhey1994,Vidal2003,calabrese2004},
including some random ones~\cite{Refael04}.
In higher dimensions, the situation is more
complicated and rigorous results are only available for the ground states of
free systems~\cite{Plenio2005,Wolf06,Gioev06} and certain classes of interacting systems.
In all of these systems, a volume law scaling will generically be observed for
high-energy eigenstates.
Here, we propose that many-body localized states have the highly
unusual property that they are excited states satisfying an area law.
As we now discuss, it is reasonable to conjecture that the area law
follows from our definition of an MBL state
and from properties of non-interacting Anderson insulators.

\begin{figure}
  \includegraphics[width=3in]{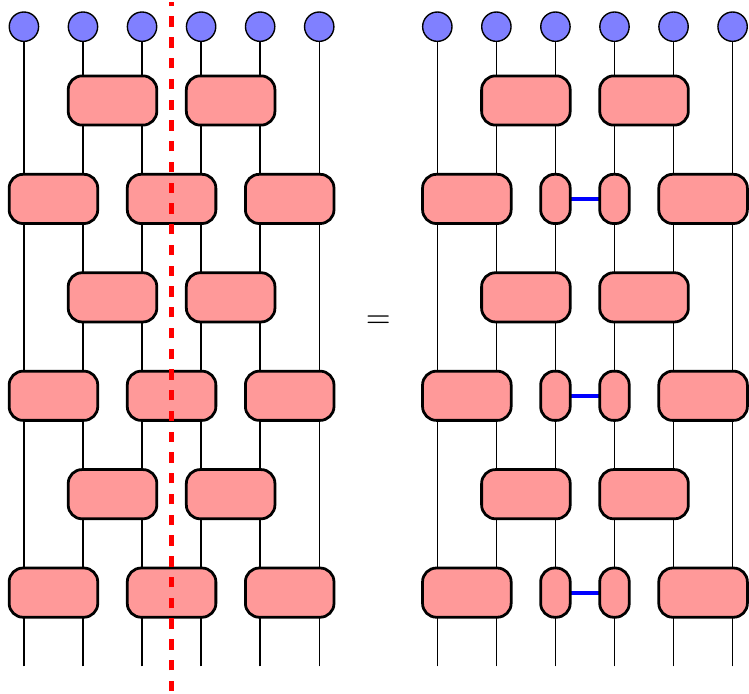}
  \caption{Transformation of a product state $|\phi\rangle$ (blue dots, top) into an entangled state by a finite-depth $2$-local (possibly unitary) transformation of depth $D=3$ (see Def.~\ref{def:U} in App.~\ref{sec:technical-details}). The amount of entanglement in the state after transformation is controlled by the number of operators that extend across a cut (such as indicated by the dashed red line on the left), i.e. by the depth of the circuit $D$. Decomposing the operators using an SVD leads to the picture on the right, which is similar to a Schmidt decomposition. The maximum entanglement is proportional to $D$. \label{fig:uc} }
\end{figure}

Consider, first, the origin of such an area law
for a system of non-interacting spinless fermions, where we fill
some set of single-particle orbitals $\phi_n(i)$. Any eigenstate of such a system can be written in the form
\begin{equation}
|\psi\rangle = \left( \sum_i \phi_1(i) c_i^\dagger \right)^{n_1} \left( \sum_i \phi_2(i) c_i^\dagger \right)^{n_2} \ldots |0\rangle,
\end{equation}
where $n_i = 0,1$ are the occupation numbers of the single-particle orbitals. Let us assume that the single-particle
orbitals $\phi_n$ are strictly local, i.e. they have support only on $m$ contiguous sites.
We can then pictorially represent
the state as a product of local operators acting on a product state, as shown in the left panel of Fig.~\ref{fig:uc}. Note that
these operators are not unitary, but this does not matter for
the purpose of our arguments here.
Using this construction, we have successfully decomposed the state $|\psi\rangle$ into a tensor network. We can now use
standard tensor network arguments to read off the maximum amount of bipartite entanglement contained in the
state $|\psi\rangle$:
Consider splitting every operator crossing the dashed red line indicated in Fig.~\ref{fig:uc} using a singular value decomposition to obtain
the picture shown in the right panel. The number of non-vanishing singular values for each operator is bounded by
$d^m$, where $d$ is the dimension of the local Hilbert space (e.g., $d=2$ for spinless fermions).
The rank of a reduced density matrix describing the left or right part is then bounded by this number to the power of
the number of operators cut, i.e. the number of blue lines in the right panel, i.e. it is bounded by $(d^m)^D$,
where $D$ is the depth of the circuit.
It follows that the bipartite entanglement entropy is bounded by
\begin{equation}
S \leq mD \log d
\end{equation}
The minimal depth $D$ of the tensor network depends on the
number of single-particle states that overlap at a given
site $i$, which is roughly $m$ and is independent of the system size.

In this argument, we have ignored the fact that we are actually interested in localized single-particle
states which can have exponential tails, rather than states with strictly local support. It is easily
shown that the overlap of a state where these tails have been truncated with the full state approaches
one exponentially quickly as the support $m$ of the strictly local states is increased, but this does
not imply that the bipartite entanglement matches to the same accuracy. However, it seems plausible
that for the eigenstates of local Hamiltonians, the entanglement properties, or at least the scaling of
entanglement entropies, does not depend on these tails for $m$ sufficiently large.

We have argued above that when the
state of the non-interacting system can be written by filling a set of strictly local
single-particle orbitals, the bipartite entanglement entropy
is independent of system size and displays an area-law scaling.
Consider now our colloquial definition for an MBL state, which is that it can be
transformed to a Slater determinant of localized single-particle states almost everywhere
to within some desired accuracy by a finite-depth local unitary circuit. Such a circuit
can, by similar arguments as given above, increase the entanglement entropy only
by some constant amount depending on its depth. By this definition, a many-body localized
state would thus inherit the area law from the corresponding Slater determinant.
An important caveat is that using a finite-depth local unitary circuit, we
can transform a state of the many-body system to a Slater determinant only to within desired accuracy
for local properties. For a rigorous argument, we would have to
show that if two eigenstates of a local Hamiltonian are locally nearly identical
then their entanglement entropies scale the same way. Although reasonable, this is a non-trivial proposition,
and we do not attempt to prove it here. We thus propose the following as a conjecture, which will be true
if small errors in the local structure of the state do not lead to errors in the scaling of the
von Neumann entropy of the reduced density matrix for a large subset of the system:
\begin{conjecture}
A many-body localized state $|\psi\rangle$ has an area law for the entanglement entropy,
that is for almost all regions of sites $\mathcal{R}$, the entanglement entropy
$S_\mathcal{R}$ has
\begin{equation} \nonumber
S_\mathcal{R} < \alpha\ \partial \mathcal{R}
\end{equation}
for some constant $\alpha > 0$.
\end{conjecture}

Here, $\partial \mathcal{R}$ denotes the boundary area of $\mathcal{R}$, and
\begin{align}
S_\mathcal{R} &=- \Tr \rho_\mathcal{R} \log \rho_\mathcal{R} &\rho_\mathcal{R} = \Tr_{\mathcal{L} \setminus \mathcal{R}} |\psi\rangle\langle\psi|.
\end{align}
This is to be contrasted with a volume law, where $S_\mathcal{R} \sim \text{Vol}\ \mathcal{R}$.
In one dimension, this means that for a system of size $L$ which is cut in a generic
location, the bipartite entanglement entropy is constant instead of proportional to $L$.
In higher dimensions, the entanglement entropy will scale as $L^{d-1}$ instead of
$L^d$.

This condition is useful for the following reason.
Although Definition \ref{def:MBL-state}
is precise, it is difficult to find the necessary local unitary transformation.
On the other hand, the entanglement entropy is more readily calculated,
given an energy eigenstate. However, our definition may still not be directly applicable
as we cannot solve the system in the thermodynamic limit, or in a sufficiently
large system to be self-averaging with enough distinct regions
${\cal R}$ of different size. In a practical calculation, we therefore have to
consider an ensemble of Hamiltonians and compare the entanglement
entropies obtained for different system sizes.

We therefore formulate the following criterion that we use to identify a
many-body localized {\it phase}. We will give a more precise definition
in Sec.~\ref{sec:MBL-phase-def}, but for now, we suggest the following
more colloquial definition:
\begin{mypar}{0.5cm}{0.5cm}
If almost all energy eigenstates of a Hamiltonian
have an area law for their entanglement entropy for almost all regions,
then we will say that the system is in a {\it many-body localized phase}.
\end{mypar}
We will make precise what we mean by "almost all"
eigenstates in Def.~\ref{def:MBL-phase}.

However, this definition is satisfied by many trivial systems,
for example systems with only on-site terms or coupling terms that are
diagonal in a product basis, such as an Ising model with a longitudinal
(instead of the transverse) field. The area law observed for arbitrary
eigenstates of such trivial systems is unstable against many perturbations,
though.
We therefore formulate the following, stricter definition which serves as sufficient and necessary condition for an MBL regime:
\begin{mypar}{0.5cm}{0.5cm}
A system described by a Hamiltonian $H$ is in a many-body localized phase
if almost all eigenstates of
\begin{equation}
H + \lambda \Phi,
\end{equation}
have an area-law for their entanglement entropy for arbitrary
perturbations $\Phi$ that are a sum of bounded local operators
and for some non-vanishing range of $\lambda$, e.g. for all $|\lambda|<\lambda_c$.
\end{mypar}

We conjecture that this definition excludes systems
that have an area law for almost
all eigenstates for trivial reasons that should not be considered many-body
localization. Thus, it paves the way for using the entanglement entropy
to identify many-body localization.

\begin{figure*}[thb]
  \includegraphics{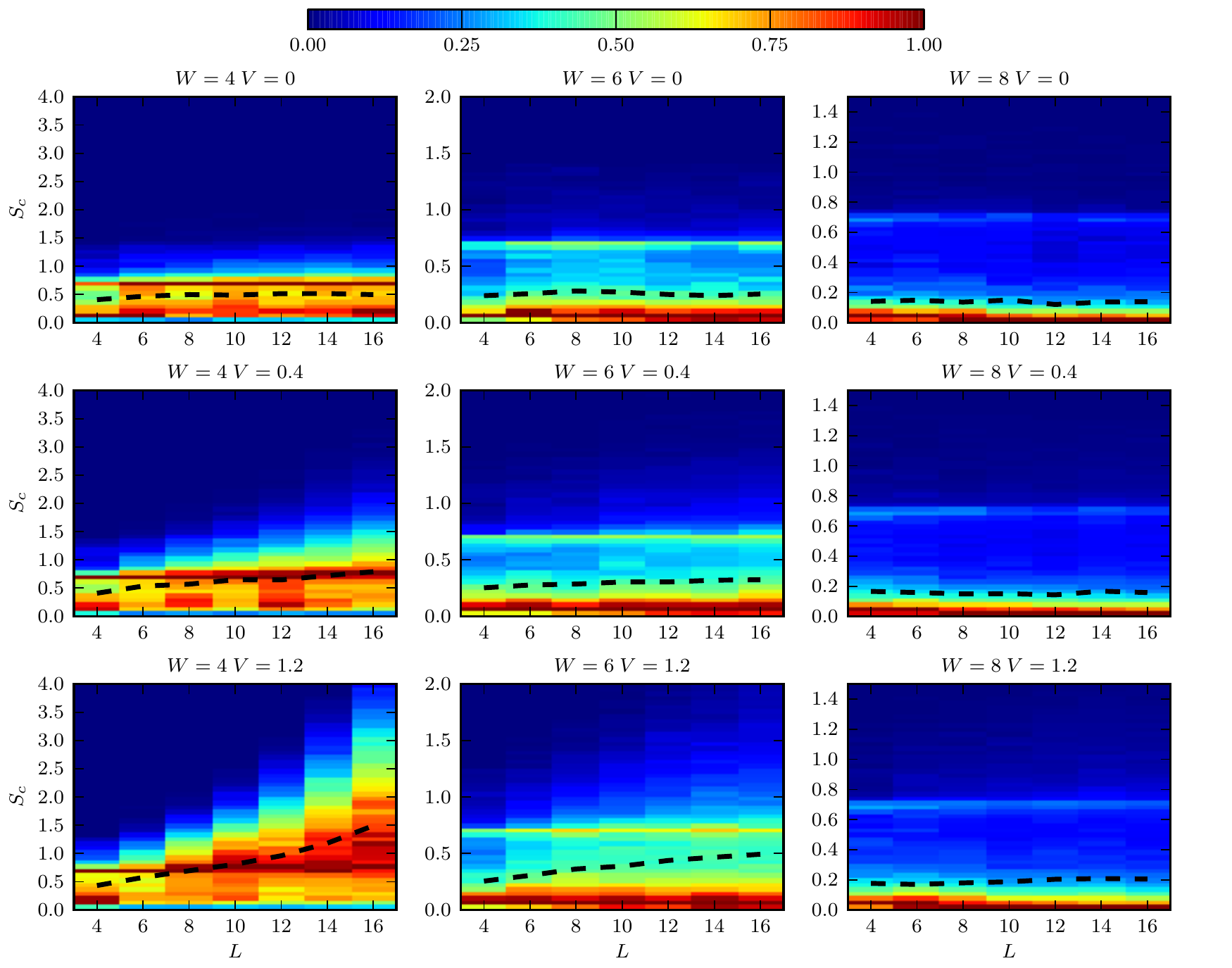}
  \caption{(Color online) Histogram of the entanglement entropy of random eigenstates of the Hamiltonian~\eqnref{eqn:H} at different values of disorder strength and interaction. Dark blue indicates a small density of states at a given $L$ and $S_c$, whereas red indicates a higher density. Dashed lines indicate the median value. Results are extracted from 1024 disorder realizations and $L$ randomly chosen states for each realization.
  \label{fig:al} }
\end{figure*}

\subsection{Numerical Calculation of Entanglement Entropy in Energy Eigenstates of
a 1D Model of Interacting Electrons in a Random Potential}
\label{sct:entropy-results}

We now turn to a numerical analysis of the entanglement entropy in random
eigenstates of the system described by Eqn.~\eqnref{eqn:H}. We calculate the
entanglement entropy at the center of the system, i.e. for a block of size $L/2$.
We perform an average over at least 1024 disorder realizations, and for each
realization extract the entropy for $L$ randomly-chosen eigenstates.
From now on, the quantity $S$ will refer to the entanglement entropy
between the left- and right-halves of the system, and we will drop the subscript
${\mathcal R}$ in $S_{\mathcal R}$ since ${\mathcal R}$ will always
be the left (or, equivalently, the right) half of the system. Occasionally,
we will use the notation $S_c$ to emphasize that this is the entropy obtained
by cutting the system at its center.

Fig.~\ref{fig:al} has histograms showing the total number of states
that have entanglement entropy $S_c$ ($y$-axis)
in the entire ensemble of 1024 systems at each size ($x$-axis)
from $L=4$ to $L=16$.
We show results for three strengths of the
disorder potential, $W=4,6,8$, and three strengths of the interaction, $V=0,0.4,1.2$. Each histogram
also shows the median value of $S_c$ (dashed line).
Consider first the non-interacting
case $V=0$ (top row in Fig.~\ref{fig:al}), where the system is a
non-interacting Anderson insulator. As expected, we observe an entanglement
entropy that is largely independent of system size but does depend on the
disorder strength since the amount of local contributions to the entanglement
entropy depends on the localization length.
For $W=6$ and $W=8$, additional peaks at $S \approx \ln(2)$ become visible.
These peaks can be traced to a purely local effect -- a localized state spread
over the two sites straddling the center of the system -- which is discussed in
Appendix~\ref{sct:log2}. However, apart from a small but noticeable number of
states that have $S \approx \ln(2)$, almost all states have entropies near or less
than the median entropy, which has very weak dependence on system size.

Although our primary interest is the scaling of $S$ with $L$, it is useful
to view $\ln(2)$ as a heuristic boundary between low- and high-entropy states.
When there are very few states with $S > \ln(2)$,
the median entropy does not scale with system size but when a large fraction
of the states of the system have $S > \ln(2)$, the median entropy tends to scale
with $L$. This is not, by any means, a precise boundary, but it is nevertheless
useful to refer to ``low-entropy states'' and ``high-entropy'' states, which
refers to their entropy relative to $\ln(2)$.

Returning to Fig.~\ref{fig:al}, we now consider the case of weak interactions
and strong disorder, $V=0.4$, $W=8$. The histogram is very similar to
those of the non-interacting case: the median entanglement entropy
depends very weakly on system size and most states have entropies
near or less than the median entropy. Even for stronger interactions,
but strong disorder, $V=1.2$, $W=8$,
the histogram is still similar to the non-interacting case, although
one might argue that a slight increase in the median $S_c$ with $L$ is visible.
But even for this strong interaction, most of the states have very low entropy.

As we decrease the disorder, we find that, for weak interactions
and moderate disorder, $V=0.4, W=6$,
there are still very few high-entropy states, and the median entropy
has weak dependence on $L$. However,
for strong interactions and moderate disorder, $V=0.4, W=6$,
the number of high-entropy states clearly increases with $L$,
and the median entropy increases with $L$ in a manner roughly consistent
with linear increase.

Finally, we consider the weak disorder case, $W=4$.
For weak interactions $V=0.4$, the number of high-entropy
states increases with system size, roughly linearly.
In this case, interactions
have caused a large fraction of the states to be
``delocalized'', i.e. to have large entanglement entropy due to
entanglement between all parts of the system.
This is reflected in the median entanglement entropy,
which grows with $L$, but slowly.  Finally, when we consider
stronger interactions, $V=1.2$,
we find that the number of high-entropy states
increases linearly with $L$ for $W=4$. The median entropy scales linearly
or, perhaps, even faster, although the apparent super-linear growth that sets
in at around $L=12$ is probably an increase in the slope from its small $L$
value to its large $L$ value.

The most salient and striking feature of these data is that
for the stronger disorder potentials $W=6,8$,
the vast majority of states at $V=0.4$ have
low entanglement entropy
for the system sizes accessible to our simulations. It seems extremely unlikely that this behavior would change for larger system sizes: if the
behavior of the system is governed by a large (or divergent)
length scale, we would expect a volume law for block sizes
smaller than this length scale and a potential crossover to an area law
beyond this scale; the other situation,
i.e. observing an area law on short scales and
a volume law at large scales, is very unlikely.
However, we cannot exclude the possibility
of extremely slow growth.

The histograms in Fig.~\ref{fig:al} include states of all energies.
However, separating the states by energy does not change the picture very much.
High entropy states occur
primarily near the center of the spectrum,
but even at the center of the spectrum there
are states with very low and very large
entropy regardless of the strength of interactions.
Hence, for the system sizes available to us,
no sharply-defined many-body analogue of a mobility edge
can be numerically observed for these
intermediate values of disorder and interaction strength.

\begin{figure}
  \includegraphics{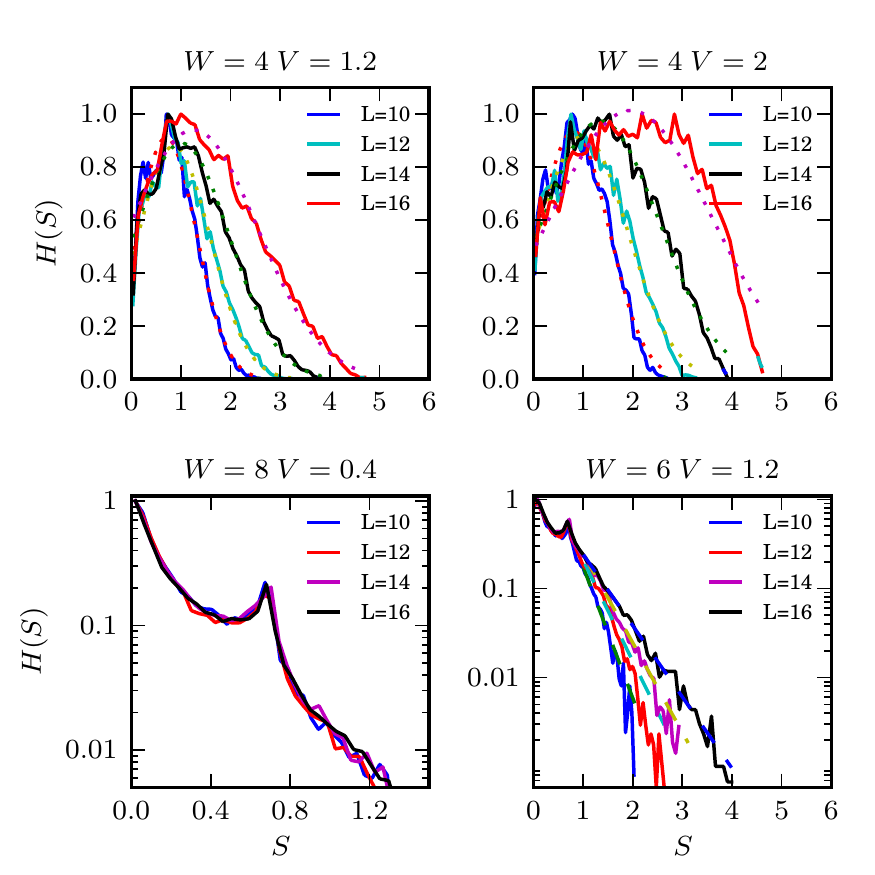}
  \caption{The number of states $H$ with entropy $S$ for different
  system sizes and system parameters.
  Plots in the top row show parameters where the system is entering a metallic
  regime and the histogram follows~\eqnref{eqn:metal}.
  In the lower left, the system is clearly in the MBL regime, where the histogram
  follows~\eqnref{eqn:aL} with $a$ independent of $L$. In the lower right panel,
  some drift of $a(L)$ with $L$ is observed, putting the system on the edge of the MBL
  regime.
  Dashed lines indicate fits to~\eqnref{eqn:aL}, dotted lines are fits to~\eqnref{eqn:metal}.
  \label{fig:tails} }
\end{figure}

\begin{figure}
  \includegraphics{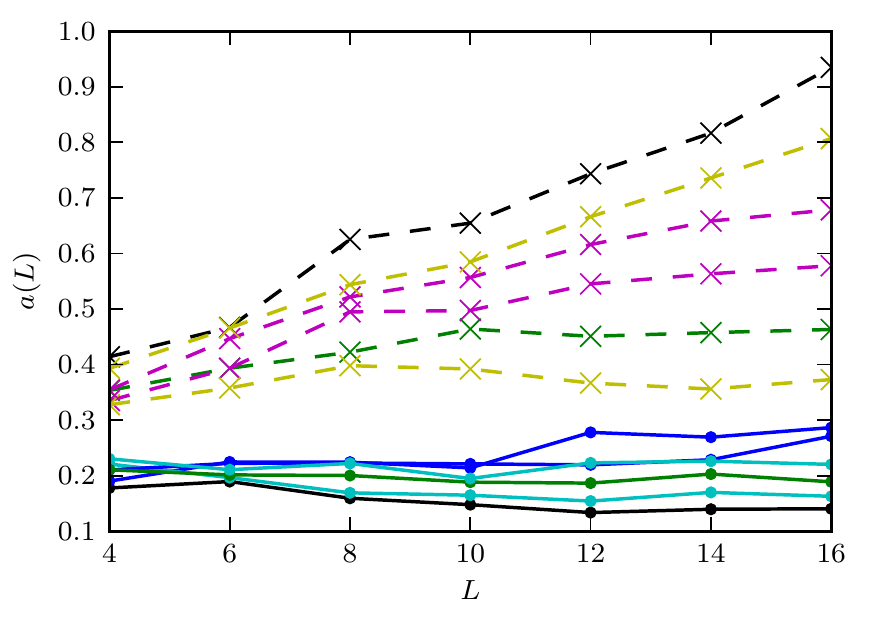}
  \caption{Coefficient $a(L)$ extracted from fits to Eqn.~\eqnref{eqn:aL}. Solid lines are $W=8$, dashed lines are $W=6$; interaction strength is $V=0,0.4,0.8,1.2,1.6,2$ from bottom to top. \label{fig:aL} }
\end{figure}


One important feature of this data, which is somewhat masked
by the color scale in Figs.~\ref{fig:al},
is that there are high entropy states at all energies. As may be seen
from Fig.~\ref{fig:tails}, these states become more rare
as we consider larger and larger entanglement entropies.
We suggest the following heuristic picture: the entanglement generally
remains small because there are ``entanglement bottlenecks''.
These are places where, for reasons which we examine in more detail
below, entanglement cannot be generated in a given state.
If we assume a probability $p_b$
that a given site is a bottleneck, then a region with entropy $S$ will
occur somewhere in the system with probability $p(S) \sim L(1-{p_b})^S$.
The probability that we will obtain this entanglement entropy from an arbitrary
cut through the system (e.g. at the middle) is $\sim (1-{p_b})^S$.
This leads to an exponentially-decaying
density $H(S)$ of states with entanglement
entropy $S$ across the mid-point of the system, $H(S) \sim e^{-S/a}$.

In Fig.~\ref{fig:tails}, we see four different types of behaviors.
From the lower left panel of Fig.~\ref{fig:tails}, we see
that, for $W=8, V=0.4$, the density of states $H(S)$
with entanglement entropy $S$ indeed
decays exponentially, as expected from the above discussion.
 (Note that there is a peak at $S=\log 2$, as discussed above.)
In the upper panels, $H(S)$ clearly does not satisfy the definition of an MBL phase
for $W=4, V=1.2$ or $2$, as we discuss below. In the lower right
panel, where $W=6, V=1.2$, the situation is unclear. This will be discussed
below.

First, however, we examine more quantitatively the extent to which the
distribution of entropies $H(S)$ agrees with our expectations. We consider
the more general form for a finite-size system:
\begin{equation} \label{eqn:aL}
H(S) \sim e^{-S/a(L)}.
\end{equation}
In a many-body localized regime, we expect $a(L)$ to approach a
finite constant as $L\rightarrow\infty$.
For $W=8$, Fig.~\ref{fig:aL} shows that $a(L)$ is
approximately constant over a wide range of
interaction strengths $V$,
as expected if the system were in an MBL phase in this range.
For $W=6$, $a(L)$ appears be constant for $V=0.4$ (second dashed line
from the bottom in Fig.~\ref{fig:aL}) but it increases
for larger values of $V$. Finally, for $W=4$ (not shown),
$a(L)$ appears to increase for all of the values of $V>0$ studied,
with the possible exception of $V=0.2$.

In a regime where the system is metallic but disorder remains strong enough to
affect the local physics, we expect that
the mean entropy scales with system size and the entropies are
normally-distributed around the mean value, so that
\begin{equation} \label{eqn:metal}
H(S) \sim e^{-{(S-{s_0}L)^2}/\alpha L}.
\end{equation}
This is observed for weak disorder, $W=4$,
and strong interactions, $V=1.2$ and $V=2$, in the upper panels of Fig.~\ref{fig:tails}.
The peak of the Gaussian is approximately at the median value of the
entropy, which increases linearly with system size $L$.
This is somewhat obscured at smaller system sizes because the $\ln(2)$
peak dominates the data. It may be somewhat surprising that
we obtain these results even without restricting to states of a single energy
because, in a metallic state in thermal equilibrium
 we expect the entropy density $S/L$ to
grow with the energy of the state.
However, the histogram is dominated by states near the center of the
(many-body) energy spectrum because there are exponentially more of them
than there are states in the tails of the spectrum. Therefore, $H(S)$ is nearly indistinguishable
from $H(S,E)$, the histogram restricted to an energy window around $E$,
if $E$ is near the center of the spectrum. On the other hand,
in the band tails $H(S,E)$ looks similar to the MBL regime.

For intermediate parameters between the MBL and
the metallic regimes, we observe a different behavior of $H(S)$.
Consider, for example, $W=6$ and $V=1.2$. From Fig.~\ref{fig:al}, we see that
for these parameters the median value of $S$ increases slowly with
$L$, consistent with $S\propto L$ but with small proportionality constant.
This is reflected in the lower right panel of Fig.~\ref{fig:tails}, where we see that
$H(S)$ decays as $H(S) \sim e^{-S/a(L)}$ with $a(L)$ increasing.
The coefficient $a(L)$  increases with system size, as may be seen in Fig.~\ref{fig:aL}. Therefore, at least for these system sizes, the system
at $W=6$ and $V=1.2$ is qualitatively different from the systems
at $W=4$ and $V=1.2$. Although the median entropy grows with system
size in both cases, the distribution of entropies is rather different.
In the latter case, it appears similar to thermodynamic equilibrium
but in the former case, it looks more like a long tail of high entropy states in an
MBL system.

In Fig.~\ref{fig:aL}, we can see that for  $W=6$ and $V=1.2$, $a(L)$ grows
linearly for small systems and the growth slows down for the largest systems
accessible to our simulations.
The simplest possibility is that the system is in an MBL
phase, and $a(L)$ approaches a finite limit for large enough systems.
An alternate possibility is that the system is in a metallic phase, but with
low $S/L$.
However, it is also possible that the system is in an intermediate phase in which
$a(L)\sim L^\beta$ at large $L$, with $0<\beta\leq 1$
so that a non-zero fraction of the states of the system
have entanglement entropy $S > \kappa L^\beta$ for some $\kappa$.
This would not be a conventional
metallic phase, even if $\beta=1$ (or, more generally, $\beta=d$ in $d$-dimensions).
Rather, it is a phase in which an $\ord{1}$ fraction of the
states has low entanglement entropy
and an $\ord{1}$ fraction has large (though possibly sub-linear)
entanglement entropy. We call this a ``long tail'' regime, in which the median entropy increases with system size -- possibly even linearly -- but the system is not in an equilibrium metallic phase because an $\ord{1}$ fraction of the
states of the system have $\ord{1}$ entanglement entropy, even
if we restrict attention to energies near the center of the spectrum.

\subsection{Definition of a many-body localized phase}
\label{sec:MBL-phase-def}

In this section, we condense the above considerations about the statistical distribution
of entanglement entropies into a concise definition for a many-body localized phase.

Let us revisit our heuristic picture for the entanglement bottlenecks. These can occur
for two reasons: There can be disorder realizations
in which the on-site potential is very large or very small at a given site,
thereby effectively cutting the system there.
Conversely, there can be disorder realizations in which the on-site potential
is nearly the same over clusters of several sites, which is analogous
to the two-site clusters that cause the $\ln(2)$ peak discussed previously
and in Appendix \ref{sct:log2}. This leads to the absence of bottlenecks
in that cluster of sites. A second reason why bottlenecks could occur
is the particular choice of state. For a fixed disorder realization, there may
be arrangements of the particles -- possibly of very high energy -- which
effectively cut the system in two, thereby forming a bottleneck.
Of course, these two mechanisms are not independent.
Regardless of the ultimate cause of the bottleneck,
the entanglement entropy across it would be low because particle and correlations
cannot propagate through a bottleneck.

\begin{figure}
  \includegraphics{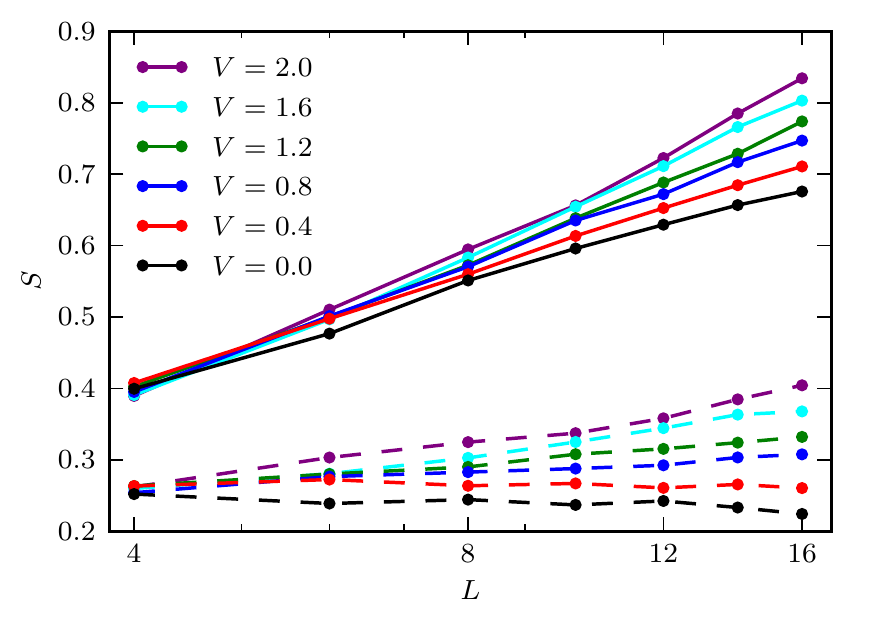}
  \caption{Disorder-averaged entropy (i) for the cut at the center of the system (dashed lines) and (ii) for the cut with the maximum entropy in each disorder realization (solid lines) for $W=6$ on a semi-logarithmic scale. While an area law is observed for the center cut for weak enough interactions, a logarithmic divergence of the entropy for the maximum entropy cut is found for these interaction strengths. For strong enough interactions, both diverge with a volume law. \label{fig:maxmean} }
\end{figure}

Let us now consider a very large system $L$, where most of the states
are MBL states. We have argued that the probability of finding entropy $S$ in such
a system for a generic cut is $H(S) \sim e^{-S/a}$;
cuts where $S/a\gg 1$ are therefore
unlikely. One could, however, also ask: what is
the probability of finding some cut of the system for which
the entanglement entropy is $S$?
This clearly follows ${\tilde H}(S) \sim Le^{-L/a}$,
and hence only cuts where $S/a \gg \ln L$ are rare; in other words, the maximum entanglement
entropy found in the system diverges logarithmically, while the median entanglement
entropy saturates to an area law. This behavior is observed in the data shown
in Fig.~\ref{fig:maxmean}. However, our arguments also suggest that for a given state of
a very large system, the distribution of entanglement entropies obtained by cutting
in different ways has a finite variance, i.e. while the maximum may diverge, the median
(and other quantiles) of the distribution remain finite. In other words, as the thermodynamic
limit is approached, the cuts that lead to a divergent entanglement entropy become a set of measure $(\ln L)/L$. Relating this back to our Def.~\ref{def:MBL-state}, we can expect that using a finite-depth
unitary transformation, an MBL state can be transformed into a Slater determinant of localized single-particle orbitals
everywhere except in an exponentially-small fraction of the system.

Note that this is very different from the case of single-particle localization.
In the single-particle case, there may be regions of size $\ln L$
in which the on-site potential is nearly
constant over the entire region. However, this is not sufficient for
a state with localization length $\sim \ln L$. As we consider larger
and larger regions, the potential must get flatter and flatter over
the entire region in order for a delocalized state to occur. Thus
states with large localization length occur only occur for disorder configurations of measure zero in the probability distribution of the on-site disorder \cite{Frohlich83}.  However, in the case of MBL, the on-site potential need not be fine-tuned to be within some set of configurations of measure zero. There could, instead, be arrangements of the particles that allow entanglement to build up
even for more generic disorder configurations.

With the preceding considerations in mind, we define a
{\it many-body localized phase} as follows.
\begin{definition}
\label{def:MBL-phase}
An infinite system described by some Hamiltonian $H$ is in an {\it MBL phase} if, for any
$f, \epsilon, k$ there exist finite real numbers $D, b, c$ such that for any $d>D$,
all energy eigenstates are MBL states with localization depth
$\leq d$ to accuracy $(\epsilon,k)$ on a volume fraction $f$ of the system,
except for a fraction of states $< b\,e^{-c d}$.
\end{definition}

Note that in a many-body localized system, generically no mobility
edge can be observed for the following reason: In a non-interacting
Anderson insulator, localized and extended single-particle states
cannot coexist at the same energy --
if there were both a localized and an extended single-particle state
at the same energy, it would require fine-tuning for them to have vanishing
matrix elements. Small changes in the Hamiltonian would mix them,
causing both states to become extended.
In the many-body system, on the other hand, low- and high-entropy states can exist
at the same energies because the matrix elements between
these many-body states could vanish due to an orthogonality catastrophe:
they may differ in the arrangements of an extensive number of particles
so that the matrix elements between them for any local Hamiltonian
would vanish.

\section{Topological Order at Non-Zero Excitation Energy Density
and Self-Correcting Quantum Memory}
\label{sct:topological}

\subsection{Overview}

A system is in a topological phase in its ground state
if it satisfies the following definition:
On a manifold ${\cal M}$, it has a set of orthonormal ground
states $|a\rangle$, $a=1,2,\ldots,{N_{\cal M}}$.
The degeneracy ${N_{\cal M}}$ depends only on the
topological configuration of the system, e.g. genus,
number of boundaries, and boundary conditions.
These states are separated from the rest of the spectrum by
an energy gap $\Delta$ that remains non-zero in the limit that the
system size $L\rightarrow\infty$.
Furthermore, for any local operator $\phi$,
\begin{equation}
\label{eqn:top-phase-def}
\langle a| \phi |b\rangle = C \delta_{ab} + \ord{e^{-L/\xi}}
\end{equation}
where $C$ is a constant independent of $a,b$;
$L$ is the system size; and $\xi$ is the correlation length of the system,
which is finite in the limit $L\rightarrow \infty$. The Hamiltonian is a sum of
local operators, so the energy splitting between the ${\cal M}$ ground
states vanishes exponentially with the system size.

In many situations, a topological phase can equivalently be defined as a gapped
phase which cannot be transformed to a local product state by a local unitary
transformation of finite depth. For a symmetry-protected topological phase, which
is not our primary focus here, one has to augment this definition by requiring that
the unitary transformation leave the symmetries untouched. However, as was shown
in Ref.~\onlinecite{Hastings11}, there is a subtle difference between these definitions
when considering disordered systems with ground states that
have a non-zero density of localized quasiparticles: these will generally not satisfy
the latter definition since the quasiparticles effectively partition the system into finite
blocks that can be transformed to a product. However, such a system could still
satisfy the first definition given above.

A similar situation is encountered for systems at non-zero temperature in two dimensions,
where a non-zero -- albeit exponentially small in $T/\Delta$ -- density of quasiparticles
is thermally induced. Such a system clearly violates the second definition and it
is not immediately clear how to extend the first definition to excited states at finite
energy density. However, it is conceivable that topological order survives even
in states of finite energy density above the ground state if the quasiparticles are
many-body localized. In this section, we seek to find a definition of a topological
phase that encompasses all these scenarios. Similar ideas were recently put
forward in Ref.~\onlinecite{Huse13}. We will largely rely on
the area law as a definition of a many-body localized regime.

\begin{figure}
  \includegraphics{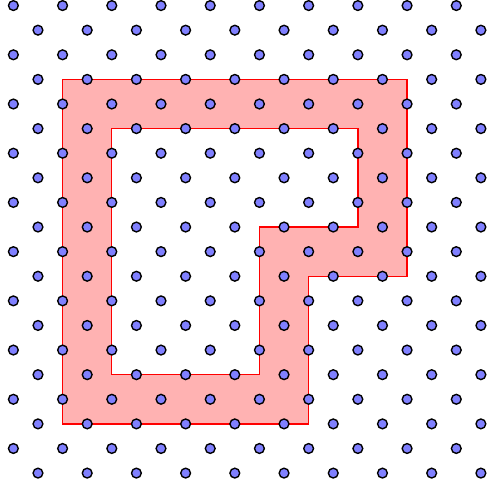}
  \caption{Fattened Wilson loop operator $W_a({\cal C})$. \label{fig:wilson} }
\end{figure}

Following Ref. \onlinecite{Hastings05}, we will define a topological
phase (in its ground state, for now) in terms of a ``zero-law'' for Wilson loop operators.
At an intuitive level, the definition is the following: We define
appropriate ``fattened'' Wilson loop operators ${W_a}({\cal C})$ associated to
curves ${\cal C}$ (cf Fig.~\ref{fig:wilson}) and quasiparticle types $a$ (see, for instance,
Ref.~\onlinecite{Nayak08} and references therein). These operators will
be defined explicitly below. In a topological phase in its ground state,
this Wilson loop operator will have the following properties:
When measured on the sphere or plane,
such operators have ground state expectation values
$\langle 0| {W_a}({\cal C})|0\rangle \neq 0$
that are independent of the perimeter $P_{\cal C}$ of ${\cal C}$.
On manifolds with non-trivial topology, such operators can be used
to distinguish the different degenerate ground states. Depending
on the curve  ${\cal C}$ and quasiparticle type $a$, Wilson loop operators
can have expectation values that are different for different ground states
or can have non-vanishing matrix elements between different ground states.
When measured in a non-topological phase, on the other hand, the expectation
values of such Wilson loop operators will decay as $e^{-aP}$ for some constant $a$.

The reason that we need to consider ``fattened'' Wilson loop operators
is that, even in the ground state of a topological phase, virtual quasiparticle
pairs will be created and one member of the pair can fluctuate between the
interior and exterior of the region bounded by ${\cal C}$. If the
Wilson loop operators are ``fattened'' so that they have a width
that is larger than the length scale for such fluctuations, then
virtual quasiparticle will not cause $\langle {W_a}({\cal C})\rangle$ to wash out.

Let us now consider the case of a topological phase at finite temperature,
i.e. at finite energy density above the ground state.
In this setting, thermally excited and mobile quasiparticles will hop in and out of the loop,
washing out $\text{Tr}\left( e^{-\beta H} {W_a}({\cal C})\right)$ and leaving behind
a strong dependence on the perimeter of the curve ${\cal C}$.
If, however, quasiparticle excitations are localized in the
energy eigenstate $|E\rangle$, this may be prevented:
Suppose that we have a topological phase with a
large gap. Then, since the gap is large, we can, with negligible effect on the ground state,
tune the Hamiltonian so that the quasiparticles have
a large effective mass and we can tune the disorder so that it has a large effect
on the quasiparticles above the gap. If quasiparticle excitations are many-body localized
in this limit, which seems ideally suited for it to occur, then excited
quasiparticles will be unable to move. Therefore, they will be unable to
wash out $\langle E| W({\cal C})|E\rangle\sim \text{const}$.

To define topological order in the ground state, we could exploit either
the ground-state degeneracy or Wilson operators. While they are
equivalent, the former is typically more convenient since it is not always
obvious how to define the appropriate Wilson loop operators.
However, as was discussed above, it is very difficult to use
degeneracy as a criterion since it may be hard to identify the correct
correspondence between degenerate states in different topological
sectors since the density of states at energies much above the spectral
gap is always exponentially large. However, as pointed out by Huse et al.\cite{Huse13},
there are scenarios where such an approach can be applied.

In essence, the Wilson construction we describe here could allow
us to identify topological
multiplets in the midst of a very high density of states.
In the case of a Hamiltonian whose clean limit
we understand, it could be possible to construct these operators
and, by entering the regime noted above, test the possibility
of topological order stabilized by many-body localization.

\subsection{Example: Toric Code}

Consider the perturbed toric code Hamiltonian~\cite{Kitaev2006} with random couplings:
\begin{multline}
  H_{\rm TC} =   -  \sum_v {J_e}(v) A(v) -  \sum_p {J_m}(p) B(p) \\
  - h \sum_i \sigma_i^z - {\tilde h} \sum_i \sigma_i^x
\label{Eq:ToricCode}
\end{multline}
Here, $A(v)$ and $B(p)$ are defined by:
\begin{equation}
A(v) \, = \!\!\prod_{j\in\mbox{\scriptsize vertex}(v)}\!\! \sigma^z_j\, , {\hskip 0.5 cm}
B(p) \, = \!\!\prod_{j\in\mbox{\scriptsize plaquette}(p)}\!\!\sigma^x_j
\end{equation}
and ${J_e}(v)$ and ${J_m}(p)$ are independent random variables chosen uniformly
from ${J_e}(v)\in \bigl[{J^e_0}-W, {J^e_0}+W\bigr]$
and ${J_m}(p)\in \bigl[{J^m_0}-{\tilde W}, {J^m_0}+{\tilde W}\bigr]$.
The magnetic fields $h$, ${\tilde h}$ create pairs of, respectively, magnetic and electric
excitations and move these excitations. In the limit that
$h, {\tilde h} \ll W,{\tilde W} \ll {J^e_0}, {J^m_0}$,
the ground state is essentially the ground
state of the toric code and the gap to all excitations is very large~\cite{Trebst2007,Hamma2008,Vidal2009,Vidal2009-1,Tupitsyn2010,Dusuel2011}.
However, at energies larger than the gap, excitations will be localized:
although $h, {\tilde h}$ could allow them to move, the disorder $W$ is
so much larger that both electric and magnetic particles will get trapped
in regions where ${J_e}(v)\approx {J^e_0}-W$ or
${J_m}(p)\approx{J^e_0}-W$, respectively. If many-body localization
occurs, then electric and magnetic particles will be unable to move,
even in states with non-zero energy densities and in the presence of
perturbations such as the following, which may be viewed as an
interaction between quasiparticles:
\begin{multline}
H_{\rm int} = V\sum_{\langle v,v'\rangle} A(v) A(v')
+ {\tilde V}\sum_{\langle p,p'\rangle} B(p) B(p').
\end{multline}
$V$ is an interaction between electric particles on neighboring vertices
while ${\tilde V}$ is an interaction between magnetic particles on
neighboring plaquettes.

For $h, {\tilde h}=0$, the Wilson loop operators
\begin{equation}
{W_m}({{\cal C}_D}) = \prod_{i\in {\cal C}_D} \sigma^z_i
\,\,\,\, , \hskip 0.4 cm
{W_e}({\cal C}) = \prod_{i\in {\cal C}} \sigma^x_i
\end{equation}
have expectation value $1$ in the ground state, i.e.
$\langle 0|{W_m}({{\cal C}_D})|0\rangle=1$,
$\langle 0|{W_e}({{\cal C}})|0\rangle=1$
for any contractible closed curves ${\cal C}$ on the lattice
and ${\cal C}_D$ on the dual  lattice. However,
for $h, {\tilde h}\neq 0$, virtual excitations of pairs of
quasiparticles straddling the curves ${\cal C}$, ${\cal C}_D$
will cause these expectation values to vanish for large loops as
$\langle 0|{W_m}({{\cal C}_D})|0\rangle \sim e^{-aP_{{\cal C}_D}}$
and $\langle 0|{W_e}({{\cal C}})|0\rangle \sim e^{-aP_{{\cal C}}}$,
where $P_{\cal C}$ is the perimeter of a the curve ${\cal C}$.
Thus, we must consider `fattened' Wilson loop operators.
Consider, for instance,
\begin{equation}
{W^F_m}({{\cal C}_D}) =  \prod_{i\in {\cal C}_D}
\frac{1}{2n+1}\sum_{j={-n}}^{n}\sigma^z_{i+j_\parallel}
\end{equation}
where $i+j_\parallel$ is the link of the lattice that is
$j$ spacings away from link $i$ in the direction of link $i$.
For $n$ much larger than the typical
separation of a virtual pair of electric excitations (which is $\sim{\tilde h}/{J^e_0}$
for small ${\tilde h}$), $\langle 0|{W^F_m}({{\cal C}_D})|0\rangle>0$
for $h>0$ but $T=0$. In other words, the topological phase survives
small quantum fluctuations at zero temperature, as can be diagnosed
with fattened Wilson loop operators.

Now consider non-zero temperature. For vanishing disorder,
$W,{\tilde W}=0$, and non-zero temperature $T>0$,
there will be a non-zero density of thermally-excited
quasiparticles. For large loops ${{\cal C}_D}$,
$\text{tr}\left(e^{-\beta H}{W^F_m}({{\cal C}_D})\right)\sim e^{aP_{{\cal C}_D}}$
because the statistical average will include nearly equal numbers
of states with even and odd numbers of quasiparticles within
each loop ${{\cal C}_D}$.

Suppose now that an MBL phase occurs in this model
in the regime $h, {\tilde h} \ll W,{\tilde W} \ll {J^e_0}, {J^m_0}$.
Then, in an MBL eigenstate of energy-density $\varepsilon$,
there will be a non-zero density $\sim\varepsilon/{J^{e,m}_0}$ of
localized quasiparticles. If this density is low and the localization
length is short, we can choose curves
${{\cal C}_D}$ and `fattened' Wilson loop operators ${W^F_m}({{\cal C}_D})$
that avoid these quasiparticles by passing between them.
(We will say a bit more about avoiding quasiparticles in the next paragraph)
Then, for any particular realization of the disorder, we have
$\langle \varepsilon |{W^F_m}({{\cal C}_D})| \varepsilon\rangle\sim \text{const.}$,
although its precise value will depend on the particular curve
and how many localized quasiparticles it encloses.
This is the analog for non-zero energy density of the situation
that occurs for two quasiparticles, which was studied by Stark {\it et al}
\cite{Stark11}. On the other hand, if we average over disorder realizations
or over locations of the curve within the system, we will obtain zero.

If we know {\it a priori} where the localized quasiparticles are,
then we can simply choose a curve that avoids them,
assuming that the density of quasiparticles $n_{\rm qp}$
satisfies $n_{\rm qp}\xi^2 \ll 1$, where $\xi$ is their localization length.
On the other hand, if we do not know where the quasiparticles
are located and simply place a curve ${\cal C}_D$ on the dual lattice
at random, then it has a probability
$\sim n_{\rm qp}\xi L_{{\cal C}_D}$ of intersecting one of these localized
quasiparticles, according to Crofton's formula.
So long as $n_{\rm qp}\xi L_{{\cal C}_D} \ll 1$,
which puts constraints on the energy density, then
there will be a range of lengths $L_{{\cal C}_D} \ll 1(n_{\rm qp}\xi)$
over which even a randomly chosen curve ${\cal C}_D$ will give
$\langle \varepsilon |{W^F_m}({{\cal C}_D})| \varepsilon\rangle$
independent of the size of the loop with high probability.

We can make this more concrete by specializing
to the case in which $J_0^e = \infty$ so that there are no electric
particles, only magnetic ones.
In a generic energy eigenstate, there will be a non-zero density
of magnetic particles. However, if they are many-body localized,
they will be essentially frozen into some locations so that a Wilson loop
operator ${W_e}({\cal C})$ can have a non-vanishing expectation value for
a curve ${\cal C}$ that avoids them. In the absence of electric particles, we can
perform a duality transformation from the toric code, which is a $\mathbb{Z}_2$
gauge theory, to the Ising model. We introduce a new set of spins
$\mu_p$ located at the centers of the plaquettes. They are defined by
\begin{eqnarray}
\mu^x_p &=& B(p) \, = \!\!\prod_{j\in\mbox{\scriptsize plaq}(p)}\!\!\sigma^x_j\cr
\mu^z_p &=& (-1)^{n_p}
\end{eqnarray}
where $n_p$ is the number of times that $p$ has been flipped,
counting from an arbitrary fixed reference state so that
$\sigma^z_i = \mu^z_p \mu^z_{p'}$ where $p$ and $p'$
share the edge $i$. The Hamiltonian now takes the form
\begin{eqnarray}
H &=& - \sum_p {J_m}(p) \mu^x_p - h \sum_{\langle p,{p'}\rangle}  \mu^z_p \mu^z_{p'}\cr
&=& - \sum_p {J_0^m} \mu^x_p - h \sum_{\langle p,{p'}\rangle}  \mu^z_p \mu^z_{p'}\cr
& &{\hskip 1 cm}  - \sum_p [{J_m}(p)-{J_0^m}] \mu^x_p
\end{eqnarray}
This is a transverse field Ising model with an average transverse
field ${J_0^m}$ and an additional random component
${J_m}(p)-{J_0^m}$ in the range $[-W,W]$.

When the transverse field Ising model is disordered, (a fattened version of)
the Wilson loop operator ${W_e}({\cal C})$ will have non-vanishing
expectation value. When it orders, the spins $\sigma^z_i$ also order,
and ${W_e}({\cal C})$ satisfies a perimeter law.
Equivalently, when the spins $\mu^z_p$ order, a domain wall
between up- and down-spins has an energy cost proportional to its length.
However, if topological order (according to our definition)
is present in some eigenstate, then the dual spins will not be ordered.
A trivial case of this occurs in the $h=0$ limit. Magnetic quasiparticles
cannot move because there is no hopping term for them.
In the dual transverse field Ising model,
there is no Ising interaction between spins, so an energy eigenstate
at finite energy density simply has a non-zero density of $\mu_p^x=-1$
dual spins. But since there are no spins fluctuating between
$\mu_p^x=1$ and $\mu_p^x=-1$, the correlation function
$\langle\mu_p^z \mu_{p'}^z\rangle$ vanishes for $p\neq p'$.
On the other hand, if $h\neq 0$ in the absence of disorder,
then the $\mu_p^x=-1$ magnetic particles can move
and, consequently, the bipartite entanglement entropy
will have a volume law corresponding to a gas of bosons.
As a result, even fattened versions of the Wilson loop
operator ${W_e}({\cal C})$ will have perimeter laws.
However, if the $\mu_p^x=-1$ magnetic particles
are many-body localized for $h\neq 0$ and $W\neq 0$,
then the bipartite entanglement entropy of the transverse
field Ising model at non-zero energy density
will have an area law.

We will call a system that realizes this scenario a topological
phase at non-zero energy density, which we will
make more precise elsewhere\cite{Bauer13}.
The basic idea is that when the situation described in the previous
paragraph holds, we can measure Wilson loop operators -- which
could be done in a real system with interferometery
experiments (see Ref.~\onlinecite{Nayak08} and references therein) -- and
thereby determine where the localized quasiparticles
are. Once this is done, we can excite further quasiparticles and braid them.
The outcome of such braiding processes will be invariant under small
deformations of the quasiparticle trajectories, namely deformations that
do not cross localized quasiparticles.

\subsection{Self-correcting quantum memories}

We now turn to the related, but distinct, problem of self-correcting quantum memories.
While the question of the stability of a topological phase at non-zero energy density
involved the characterization of eigenstates of the system above the spectral gap, the
question of stability of self-correcting quantum memories relies on a characterization
of the dynamics of the system at long times. The effect of disorder on the stability of
quantum memories has previously been discussed for the toric code\cite{Wootton2011,Stark11}
and a one-dimensional Majorana chain\cite{Bravyi12}, and it was found that disorder can, under
the right circumstances, improve the stability of quantum memories.
To pose the question, let us suppose that we have quantum
information encoded in a physical system. We shield the system as well as we can, but
there will always be some rate, however small, at which external perturbations such as
cosmic rays will pass through our
shielding and cause a transition of the system into some other state. If the system
is self-correcting, then it means that the
quantum information encoded in the state -- but not necessarily other properties of the state --
is unaffected at long times; more precisely, we will require the error rate to be exponentially-small
in the system size.

Consider, for the sake of concreteness, the Toric code in four
dimensions\cite{Dennis2002},
which is an example of a topological phase that survives
at $T>0$\cite{Alicki2010}. Its stability at $T>0$ is due to the
fact that the quasiparticles are string-like excitations which have a string tension.
Consequently, an energy proportional to the linear size of the system
is required to create a long-enough string to
change the topological sector of the system.
Let us suppose that the system is initially in some state
in the $N_{\cal M}$-dimensional ground state subspace of the full Hilbert space.
Every time a high-energy photon enters the system, there is some probability
that it creates a pair of small string loop excitations.
The system, which is in a higher-energy state in the same topological sector,
can now relax back to the ground state subspace by
pair-annihilating the small string loops and emitting a photon.
In order for an error to occur, one of the string loops must wrap around the system
before they annihilate. However, it costs energy to increase the length
of a string loop. This energy must be supplied by another photon.
If the rate at which photons enter the system is sufficiently small, then
the system will almost surely relax back before the string can grow.
There is a very small probability $p$ for the string loop to
grow larger as a result of a second photon before
the error has been relaxed away; the probability that a series
of such events allows a string to grow to length $L$ is $p^L$. Therefore,
if this scenario is correct, then the probability that an error occurs is
exponentially-small in the system size, i.e. the system is a self-correcting quantum memory.

Consider, in contrast, a clean topological phase in two
dimensions\cite{Alicki2008}, which is known to be unstable
to any non-zero temperature. Suppose an incoming photon is absorbed by the system
and a quasiparticle-quasihole pair is created. Then, the quasiparticle-quasihole pair
could annihilate, emitting a photon. But even if the probability for this is large, there will
be some non-zero probability for the quasiparticle and quasihole to move apart from
each other rather than immediately annihilate. Since there is no long-distance force
holding the quasiparticle and quasihole together, once they move apart, they
can move around independently. Eventually, they will meet again and
annihilate, but the probability for the difference between the
quasiparticle and quasihole trajectories to be a loop
encircling the system decreases as a power of the system size $L$.
Therefore, even at $T=0$, just a single incoming photon will
cause an error with a probability that decreases as a power of the system size
(see, e.g., Ref.~\onlinecite{Alicki2006}).

Now suppose that the system is dirty and, furthermore,
that all quasiparticle and quasihole states
are many-body localized. A single incoming photon can create a quasiparticle-quasihole pair,
but the quasiparticle and quasihole cannot move on their own and
eventually the pair will annihilate and emit a photon.
Even if this takes a long time, a single photon
cannot lead to an error since the quasiparticle and quasihole can't move sufficiently
far away to change the information encoded in the quantum state, and hence.
If, on the other hand, a second photon impinges on
the system before the quasiparticle-quasihole pair annihilates,
it could provide one of them with
sufficient energy to hop into another localized state.
Provided that an external source pumps energy into the system at a constant
and sufficiently high rate,
the quasiparticle and quasihole may
execute a random walk and move far apart.
In a finite system, they will eventually meet again and annihilate,
but there will be a non-zero probability
for the difference between the quasiparticle and quasihole trajectories to be a loop
which encircles the system, leading to a change of the information encoded in the system.
To summarize, although a single incoming photon cannot cause an error,
a non-zero rate of incoming photons can cause errors
with probability inversely proportional to a power of $L$ even in a many-body
localized system. Since this is not an exponential, it is not a self-correcting
quantum memory.

\section{Discussion}
\label{sct:discussion}

A disorder-driven metal-insulator transition is, seemingly,
a transition in the purely dynamical properties
of a system. However, \emph{single-particle localization} can be diagnosed
from the properties of an individual wavefunction. For instance,
the single-particle eigenstate at a given energy can be computed by the transfer
matrix method, and the properties of the state can be deduced from
the eigenvalues of the transfer matrix -- the system is localized
when there is a gap between the two lowest eigenvalues.
In this paper, we suggest a definition of \emph{many-body localization}
in terms of individual energy eigenstates. We define a many-body
localized state as one which can be transformed almost everywhere
into a Slater determinant
of localized single-particle states by finite-depth local unitary transformations,
to within desired accuracy.

One important potential consequence, which we conjecture to follow
from our definition of an MBL state, is that
the entanglement entropy of a block
within the system scales as the surface area of that block. A many-body localized
system hence has the highly unusual property that nearly all eigenstates display
an area law.
Equivalently, the interior of a region is only correlated with the
exterior through local correlations at the boundary.

We find clear evidence for two regimes in a 1D system of
spinless fermions in a random on-site potential. For weak interactions,
the entanglement entropy is independent of the system size
(which is an area law in a 1D system). For sufficiently large
interactions, the entanglement entropy is proportional to the system size.
The former is not quite a proof that the system is many-body localized,
according to our definition. However, the other known examples
of states with area-laws {\it in excited states}
are not likely to be realized in this context.
Therefore, we take our results, summarized in Fig.~\ref{fig:al}, as a strong
indication that there is a weak-interaction regime in which the system's
eigenstates are many-body localized such that they can be transformed
into a Slater determinant of localized single-particle states
by a finite-depth local unitary transformation.

Our physical picture is that, in an MBL phase, entanglement bottlenecks emerge,
which prevent an extensive entanglement entropy from developing.
This leads to an area law but, assuming that the probability
for an entanglement bottleneck to occur at a given site is less than $1$,
it also implies that there will be an exponentially-decaying density of states with high entropies even in the regime that we identify as the MBL phase.
Our data is consistent with this and further suggests the interesting possibility of
a regime in which the median entropy increases with system size as a result of ever lengthening tails in $H(S)$, as in the lower right panel of
Fig.~\ref{fig:tails} and the upper curves in Fig.~\ref{fig:aL}.
This implies a broadening distribution of entropies, rather than a narrow
distribution about an increasing mean, as we find for the metallic states in
the upper panels of Fig.~\ref{fig:tails}.

Our definition and our calculations focus on the entanglement entropy
of energy eigenstates. This shows very different behavior than the
entanglement entropy that is generated by the dynamics of a
system that is prepared in an initial local product state and then subjected
to time evolution under the full, possibly interacting, Hamiltonian.
For clean systems, it has been shown that the entanglement entropy
can at most grow linearly in time\cite{Bravyi2006,Eisert2010} and
approaches a volume law for long times.
In contrast to this, Burrell and Osborne\cite{Burrell2007} have shown
that the entanglement growth in a non-interacting, localized system can
be at most logarithmic with time.
Several authors\cite{Znidaric08,DeChiara2006,Bardason12,Vosk13} have studied the
entanglement growth for interacting disordered systems
numerically or using a dynamical real-space renormalization approach
and have found that the entropy grows logarithmically after an initial delay,
but ultimately approaches a volume law at large enough times.
Hence, for both a localized and a delocalized system, the entropy approaches
a volume law at very long times, but only the saturation value is different between
the two cases: in the localized phase, it is smaller than in an ergodic phase.
The entanglement entropy of an eigenstate, meanwhile, appears
to show a very dramatic difference between many-body localized regime
and a metallic one: the entanglement entropy satisfies an area law in
one case and a volume law in the other.

According to our definition, if we look at a subsystem of a large system,
it does not look like a system in thermal equilibrium. For instance, its
entropy will not be extensive,
which is an indication that it is not exploring all of the available states
at that energy. This also has dramatic consequences for topological
order in highly-excited states. If excited quasiparticles -- which
would ordinarily destroy topological order -- cannot move,
then one might expect that topological order can survive.
In this manuscript, we have defined topological order at finite temperature
in terms of the expectation values of `fattened'
Wilson loop operators \cite{Levin06}. Such an operator can
avoid localized quasiparticles and give a constant determined by
the number of quasiparticles enclosed, rather
than a perimeter law. A 2D system at non-zero energy density
exhibiting topological order
according to this definition would share many properties with the ground state
of a topological phase in a disordered system
with a non-zero density of localized quasiparticles,
but it does not give a self-correcting quantum memory.

\acknowledgments
We thank David Huse for his comments on an earlier draft of this paper
and his careful explanation of his recent work.
We also have the pleasure of acknowledging useful discussions with Joel Moore, Matt Hastings,
Michael Freedman, Bryan Clark, Kirill Shtengel, and Joseph Rudnick.
C.N. has been partially supported by the DARPA QuEST
program and AFOSR under grant FA9550-10-1-0524.
Simulations were performed using the ALPS libraries~\cite{bauer2011-alps}.

During the final stages of the preparation of this manuscript, we became
aware of Refs.~\onlinecite{Huse2013,Serbyn2013}, where related ideas
are discussed.

\appendix
\section{Localized Single-Particle States and Finite-Depth
Unitary Transformations}
\label{sec:technical-details}

\subsection{Localized Single-Particle States}

In Section \ref{sct:entropy} we have made extensive use
of `localized single-particle states'; here, we give them a precise
definition.
By `localized single-particle states', we mean normalized single-particle
wavefunctions $\phi(x)$ that decay exponentially at long distances.
This can be made more precise by demanding that if
$\phi(x)$ is a localized single-particle state then
for any $\epsilon>0$, there is a function $\chi(x)$
that satisfies the following conditions.
(1) $\chi(x)$ is normalized:
\begin{equation}
\left|\int {d^d}x \,{\chi^*}(x) \chi(x) \right| =1\, .
\end{equation}
(2) The overlap between $\chi(x)$ and $\phi(x)$ satisfies:
\begin{equation}
\left|\int {d^d}x \, {\chi^*}(x) \phi(x) \right| > 1 - \epsilon \, .
\end{equation}
(3) $\chi(x)$ has compact support ${\cal R}$ of linear size
$L_{\cal R}(\epsilon)$ satisfying
$\lim_{\epsilon\rightarrow 0} [{\epsilon^\alpha} L_{\cal R}(\epsilon)]=0$
for all $\alpha>0$.

We now give a definition of a finite-depth local unitary circuit:
\begin{definition} \label{def:U}
An {\it $m$-local unitary circuit of depth $D$} is a unitary operator
\begin{equation} \nonumber
U = \prod_{i=1}^D \prod_{j=0}^{m-1} \left( U^i_j \otimes U^i_{m+j} \otimes U^i_{2m+j} \otimes \ldots \right),
\end{equation}
where $U^i_n$ is a unitary operator that acts on at most $m$ consecutive sites starting at site $n$.
\end{definition}

\section{$\log(2)$ peak in $H(S)$}
\label{sct:log2}

\begin{figure}
  \includegraphics{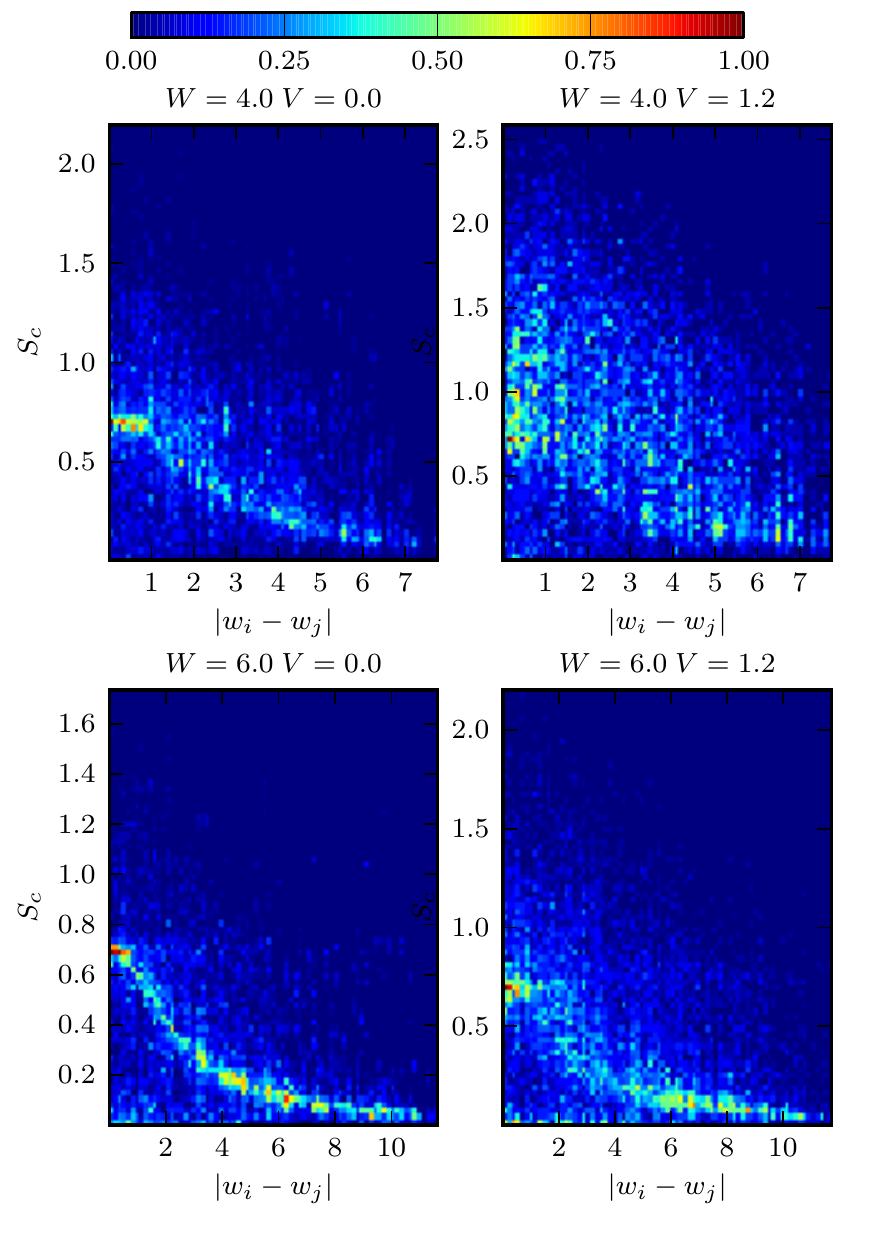}
  \caption{(Color online) Histogram of the entropy at the center of the system $S_c$ vs the difference in on-site potential between the two sites $i$, $j$ adjacent to the cut. Darker red colors indicate peaks in the histogram. The figure demonstrates that the additional peak at $S_c = \log 2$ observed in the entropy data is due to adjacent sites having very close on-site potential and thus having an electron resonate between the two sites. \label{fig:log2} }
\end{figure}

To investigate the nature of the additional peaks observed in the histogram $H(S)$ at $S \approx \log 2$,
we have plotted in Fig.~\ref{fig:log2} histograms of the entanglement entropy
at the center of the system against the difference between the on-site potentials
adjacent to the central bond, $|w_i - w_j|$.
The data clearly shows that configurations with $S_c \approx \log 2$ occur
predominantly when the difference $|w_i - w_j|$ is very small.
We conclude from this correlation that the peaks at $S_c \approx \log 2$ are an artifact
due to configurations that favor electrons
at the center to spread over two sites adjacent to the central cut.

\section{Localization in Fock space}
\label{sct:fock-app}

\begin{figure}
  \includegraphics[width=3.3in]{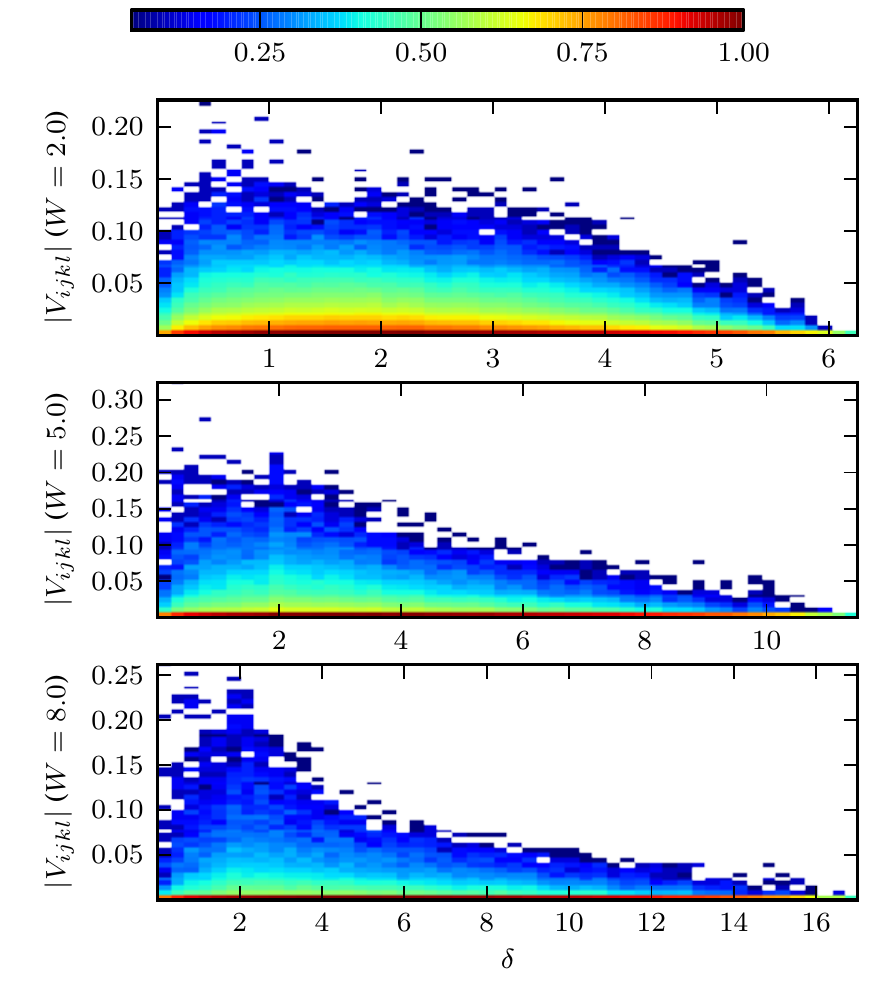}
  \caption{Normalized logarithm of the histogram, $\ln H$, of the strength of four-fermion terms $|V_{ijkl}|$ vs.
  the energy difference $\delta$ of the single-particle orbitals connected by this interaction term, cf.~\eqnref{eqn:delta}. Results are for $L=24$ and $t=V=1$ and accumulated over 100 disorder realizations. White regions correspond to $H=0$. \label{fig:V} }
\end{figure}

\begin{figure}
  \includegraphics{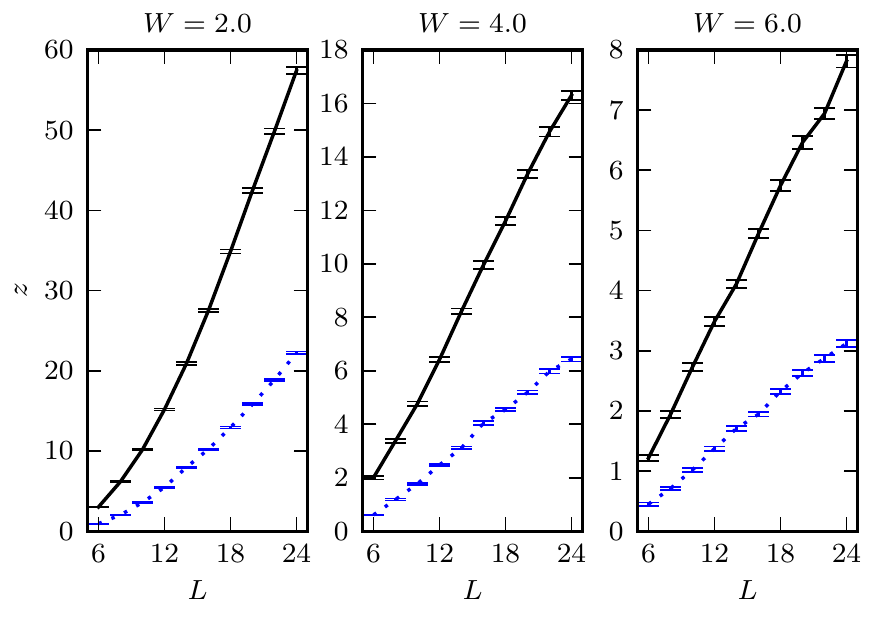}
  \caption{Scaling of the effective coordination number $z$ of Eqn.~\eqnref{eqn:z} with system size $L$ at half filling $N=L/2$ for $V=1.2$ (black, solid lines) and $V=0.4$ (blue, dotted lines). \label{fig:x} }
\end{figure}

\begin{figure}
  \includegraphics{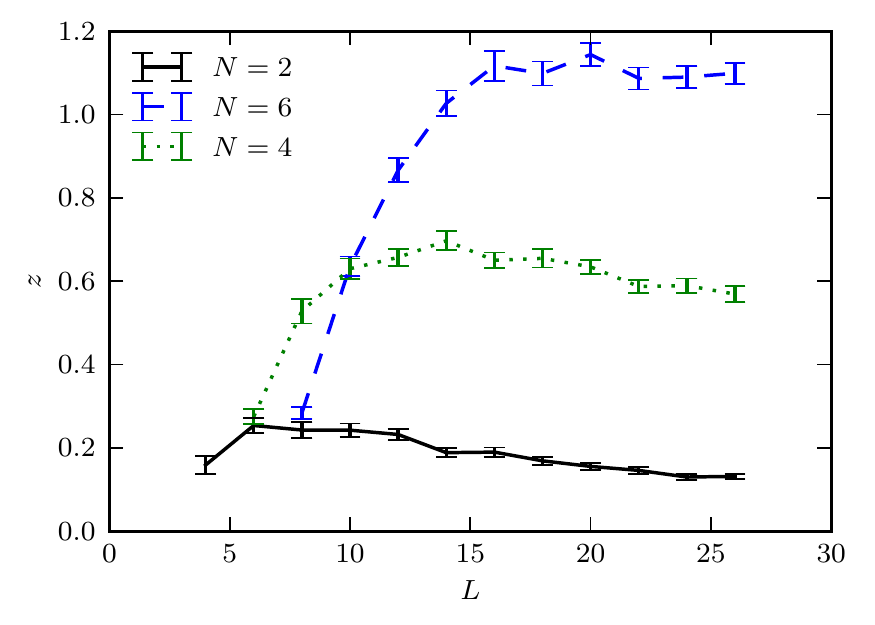}
  \caption{Scaling of the effective coordination number $z$ of Eqn.~\eqnref{eqn:z} with system size $L$ at particle number $N$, for $N=2,4,6$. Calculations are performed at $W=8$ and $V=0.4$, i.e. in a strongly localized regime. \label{fig:xn} }
\end{figure}

In this section, we will discuss some numerical results on the Fock space localization problem introduced in Section~\ref{sct:fock}. For sake of readability, we repeat the derivation of Eqn.~\eqnref{eqn:Hfock} here, giving some additional details. We begin by solving the non-interacting part $H_0$ of Hamiltonian~\eqnref{eqn:H} to obtain the eigenvalues $\varepsilon_n$ and eigenvectors $\phi_n(i)$. We can then rewrite the problem as follows (cf.~\eqnref{eqn:Hsp}):
\begin{equation}
H = \sum \varepsilon_n d_n^\dagger d_n + \sum_{ijkl} V_{ijkl} d_i^\dagger d_j^\dagger d_k d_l,
\end{equation}
where $d_n^\dagger$ creates a fermion in the single-particle state $\phi_n(i)$, and the coefficients of the four-fermion term are
\begin{equation}
V_{ijkl} = V \sum_{p=1}^{L-1} \phi_i^*(p) \phi_j^*(p+1) \phi_k(p+1) \phi_l(p).
\end{equation}

This defines a Fock space model, in which we interpret the Slater
determinants formed by filling single-particle eigenstates as sites of our new model.
We denote such a Slater determinant as
$|\vec{\alpha}\rangle = \prod (d_n^\dagger)^{\alpha_n} |0\rangle$, where $\vec{\alpha}$ is an occupation-number vector and $d_n^\dagger = \sum_i \phi_n(i) c_i^\dagger$.  There are $2^L$ such sites in the full Fock space; restricting to half filling, we are left with $\binom{L}{L/2}$ sites. We thus obtain the Fock space model (cf.~\eqnref{eqn:Hfock})
\begin{subequations} \label{eqn:Hfock-app} \begin{align}
H &= \sum_{\vec{\alpha}} \mu_{\alpha} |\vec{\alpha}\rangle \langle \vec{\alpha} | +\sum_{\vec{\alpha}\neq\vec{\beta}} V_{\beta \alpha} |\vec{\beta}\rangle\langle\vec{\alpha}| \\
\mu_{\alpha} &= \sum_n \alpha_n \varepsilon_n + V_{\alpha \alpha} \\
V_{\beta \alpha} &= \sum_{ijkl} V_{ijkl} \langle \vec{\beta} | d_i^\dagger d_j^\dagger d_k d_l | \vec{\alpha} \rangle.
\end{align} \end{subequations}

We will now numerically analyze the structure of the effective hopping terms
$V_{\alpha \beta}$. Since we are assuming that we only have two-body interactions,
the many-body states $|\alpha\rangle$ and $|\beta\rangle$ can only differ
in the occupations of four of the single particle states $\phi_i$,
$\phi_j$, $\phi_k$, $\phi_l$, which directly relates the $V_{\alpha \beta}$ to the $V_{ijkl}$.
As discussed previously, due to the localization of single-particle states $\phi_n(i)$, the
matrix elements $V_{ijkl}$ fall off exponentially with separation between the states. We now
want to confirm numerically that they also fall off exponentially with the difference in
single-particle energies.

To this end, let us consider only terms where $i \neq j \neq k \neq l$ and define
\begin{align} \label{eqn:delta}
\delta = \min \lbrace
  &\max(|\varepsilon_i - \varepsilon_k|, |\varepsilon_j - \varepsilon_l|), \nonumber \\
  &\max(|\varepsilon_i - \varepsilon_l|, |\varepsilon_j - \varepsilon_k|)
 \rbrace.
\end{align}
This defines a natural measure for the energy difference of the single-particle eigenstates involved in the term $V_{ijkl}$. In Fig.~\ref{fig:V}, we show a histogram of how the strength of $|V_{ijkl}|$ depends on this distance $\delta$. Following \baa, we expect that terms with large $\delta$ are strongly suppressed in a sufficiently strongly disordered system. This is nicely confirmed in our data: for large enough $\delta$, only very weak terms $|V_{ijkl}| \ll 1$ occur, whereas for small $\delta$ terms of all strength up to $|V_{ijkl}|  \sim \ord{1}$ occur. Note that for any choice of disorder strength, most of the terms are very small. For terms of the form $V_{ikkj}$, i.e. where only one fermions hops, very similar results are obtained with $\delta = | \varepsilon_i - \varepsilon_j |$; the probability of having terms of strength $\ord{1}$ in this case is independent of $|\varepsilon_i - \varepsilon_k|$ and $|\varepsilon_j - \varepsilon_k|$. Terms of the form $V_{ijij}$, which are diagonal, have a strength that is largely independent of $|\varepsilon_i - \varepsilon_j|$.

Turning our attention to the structure of the hopping problem of~\eqnref{eqn:Hfock-app}, we now examine the effective coordination number previously defined in~\eqnref{eqn:z}:
\begin{align} \label{eqn:z-app}
z&=\left \langle z_\alpha \right\rangle, &z_\alpha &= \sum_{\beta\neq\alpha} \frac{V_{\beta \alpha}}{|\mu_\beta - \mu_\alpha|}
\end{align}
Due to practical considerations, we only evaluate terms with $|V_{ijkl}| > \epsilon$ with $\epsilon = 10^{-3}$; our estimate for $z$ can thus be considered a lower bound, but we have confirmed that the estimate is independent of $\epsilon$ within statistical error bars. Our results for a half-filled system are shown in Fig.~\ref{fig:x}, which shows the scaling approaches $z \sim L$ largely independently of the parameters of the model. To obtain a better understanding, we study the dependence of $z$ on $L$ at fixed particle number instead of fixed filling fraction. We show $z$ as a function of $L$ in Fig.~\ref{fig:xn} for system sizes up to $L=26$ and fixed particle number $N=2,4,6$. In all three cases, a saturation of $z$ is observed for $L = cN$, where $c$ is a constant which is roughly $c \approx 3$ for our choice of parameters. This coefficient will depend on the single-particle localization length and therefore on the disorder strength $W$.

\bibliography{mbl}

\begin{thebibliography}{69}%
\makeatletter
\providecommand \@ifxundefined [1]{%
 \@ifx{#1\undefined}
}%
\providecommand \@ifnum [1]{%
 \ifnum #1\expandafter \@firstoftwo
 \else \expandafter \@secondoftwo
 \fi
}%
\providecommand \@ifx [1]{%
 \ifx #1\expandafter \@firstoftwo
 \else \expandafter \@secondoftwo
 \fi
}%
\providecommand \natexlab [1]{#1}%
\providecommand \enquote  [1]{``#1''}%
\providecommand \bibnamefont  [1]{#1}%
\providecommand \bibfnamefont [1]{#1}%
\providecommand \citenamefont [1]{#1}%
\providecommand \href@noop [0]{\@secondoftwo}%
\providecommand \href [0]{\begingroup \@sanitize@url \@href}%
\providecommand \@href[1]{\@@startlink{#1}\@@href}%
\providecommand \@@href[1]{\endgroup#1\@@endlink}%
\providecommand \@sanitize@url [0]{\catcode `\\12\catcode `\$12\catcode
  `\&12\catcode `\#12\catcode `\^12\catcode `\_12\catcode `\%12\relax}%
\providecommand \@@startlink[1]{}%
\providecommand \@@endlink[0]{}%
\providecommand \url  [0]{\begingroup\@sanitize@url \@url }%
\providecommand \@url [1]{\endgroup\@href {#1}{\urlprefix }}%
\providecommand \urlprefix  [0]{URL }%
\providecommand \Eprint [0]{\href }%
\providecommand \doibase [0]{http://dx.doi.org/}%
\providecommand \selectlanguage [0]{\@gobble}%
\providecommand \bibinfo  [0]{\@secondoftwo}%
\providecommand \bibfield  [0]{\@secondoftwo}%
\providecommand \translation [1]{[#1]}%
\providecommand \BibitemOpen [0]{}%
\providecommand \bibitemStop [0]{}%
\providecommand \bibitemNoStop [0]{.\EOS\space}%
\providecommand \EOS [0]{\spacefactor3000\relax}%
\providecommand \BibitemShut  [1]{\csname bibitem#1\endcsname}%
\let\auto@bib@innerbib\@empty
\bibitem [{\citenamefont {{Basko}}\ \emph {et~al.}(2006)\citenamefont
  {{Basko}}, \citenamefont {{Aleiner}},\ and\ \citenamefont
  {{Altshuler}}}]{Basko06a}%
  \BibitemOpen
  \bibfield  {author} {\bibinfo {author} {\bibfnamefont {D.~M.}\ \bibnamefont
  {{Basko}}}, \bibinfo {author} {\bibfnamefont {I.~L.}\ \bibnamefont
  {{Aleiner}}}, \ and\ \bibinfo {author} {\bibfnamefont {B.~L.}\ \bibnamefont
  {{Altshuler}}},\ }\href {\doibase 10.1016/j.aop.2005.11.014} {\bibfield
  {journal} {\bibinfo  {journal} {Annals of Physics}\ }\textbf {\bibinfo
  {volume} {321}},\ \bibinfo {pages} {1126} (\bibinfo {year}
  {2006})}\BibitemShut {NoStop}%
\bibitem [{\citenamefont {{Basko}}\ \emph {et~al.}()\citenamefont {{Basko}},
  \citenamefont {{Aleiner}},\ and\ \citenamefont {{Altshuler}}}]{Basko06b}%
  \BibitemOpen
  \bibfield  {author} {\bibinfo {author} {\bibfnamefont {D.~M.}\ \bibnamefont
  {{Basko}}}, \bibinfo {author} {\bibfnamefont {I.~L.}\ \bibnamefont
  {{Aleiner}}}, \ and\ \bibinfo {author} {\bibfnamefont {B.~L.}\ \bibnamefont
  {{Altshuler}}},\ }\href@noop {} {\enquote {\bibinfo {title} {{On the problem
  of many-body localization}},}\ }\Eprint
  {http://arxiv.org/abs/cond-mat/0602510} {arXiv:cond-mat/0602510} \BibitemShut
  {NoStop}%
\bibitem [{\citenamefont {Anderson}(1958)}]{Anderson58}%
  \BibitemOpen
  \bibfield  {author} {\bibinfo {author} {\bibfnamefont {P.~W.}\ \bibnamefont
  {Anderson}},\ }\href {\doibase 10.1103/PhysRev.109.1492} {\bibfield
  {journal} {\bibinfo  {journal} {Phys. Rev.}\ }\textbf {\bibinfo {volume}
  {109}},\ \bibinfo {pages} {1492} (\bibinfo {year} {1958})}\BibitemShut
  {NoStop}%
\bibitem [{\citenamefont {{Basko}}\ \emph {et~al.}(2007)\citenamefont
  {{Basko}}, \citenamefont {{Aleiner}},\ and\ \citenamefont
  {{Altshuler}}}]{Basko07}%
  \BibitemOpen
  \bibfield  {author} {\bibinfo {author} {\bibfnamefont {D.~M.}\ \bibnamefont
  {{Basko}}}, \bibinfo {author} {\bibfnamefont {I.~L.}\ \bibnamefont
  {{Aleiner}}}, \ and\ \bibinfo {author} {\bibfnamefont {B.~L.}\ \bibnamefont
  {{Altshuler}}},\ }\href {\doibase 10.1103/PhysRevB.76.052203} {\bibfield
  {journal} {\bibinfo  {journal} {\prb}\ }\textbf {\bibinfo {volume} {76}},\
  \bibinfo {pages} {052203} (\bibinfo {year} {2007})}\BibitemShut {NoStop}%
\bibitem [{\citenamefont {Deutsch}(1991)}]{Deutsch91}%
  \BibitemOpen
  \bibfield  {author} {\bibinfo {author} {\bibfnamefont {J.~M.}\ \bibnamefont
  {Deutsch}},\ }\href {\doibase 10.1103/PhysRevA.43.2046} {\bibfield  {journal}
  {\bibinfo  {journal} {Phys. Rev. A}\ }\textbf {\bibinfo {volume} {43}},\
  \bibinfo {pages} {2046} (\bibinfo {year} {1991})}\BibitemShut {NoStop}%
\bibitem [{\citenamefont {{Srednicki}}(1994)}]{Srednicki94}%
  \BibitemOpen
  \bibfield  {author} {\bibinfo {author} {\bibfnamefont {M.}~\bibnamefont
  {{Srednicki}}},\ }\href {\doibase 10.1103/PhysRevE.50.888} {\bibfield
  {journal} {\bibinfo  {journal} {\pre}\ }\textbf {\bibinfo {volume} {50}},\
  \bibinfo {pages} {888} (\bibinfo {year} {1994})}\BibitemShut {NoStop}%
\bibitem [{\citenamefont {Altshuler}\ \emph {et~al.}(1997)\citenamefont
  {Altshuler}, \citenamefont {Gefen}, \citenamefont {Kamenev},\ and\
  \citenamefont {Levitov}}]{Altshuler1997}%
  \BibitemOpen
  \bibfield  {author} {\bibinfo {author} {\bibfnamefont {B.~L.}\ \bibnamefont
  {Altshuler}}, \bibinfo {author} {\bibfnamefont {Y.}~\bibnamefont {Gefen}},
  \bibinfo {author} {\bibfnamefont {A.}~\bibnamefont {Kamenev}}, \ and\
  \bibinfo {author} {\bibfnamefont {L.~S.}\ \bibnamefont {Levitov}},\ }\href
  {\doibase 10.1103/PhysRevLett.78.2803} {\bibfield  {journal} {\bibinfo
  {journal} {Phys. Rev. Lett.}\ }\textbf {\bibinfo {volume} {78}},\ \bibinfo
  {pages} {2803} (\bibinfo {year} {1997})}\BibitemShut {NoStop}%
\bibitem [{\citenamefont {Berkovits}\ and\ \citenamefont
  {Avishai}(1998)}]{Berkovits1998}%
  \BibitemOpen
  \bibfield  {author} {\bibinfo {author} {\bibfnamefont {R.}~\bibnamefont
  {Berkovits}}\ and\ \bibinfo {author} {\bibfnamefont {Y.}~\bibnamefont
  {Avishai}},\ }\href {\doibase 10.1103/PhysRevLett.80.568} {\bibfield
  {journal} {\bibinfo  {journal} {Phys. Rev. Lett.}\ }\textbf {\bibinfo
  {volume} {80}},\ \bibinfo {pages} {568} (\bibinfo {year} {1998})}\BibitemShut
  {NoStop}%
\bibitem [{\citenamefont {Mej\'ia-Monasterio}\ \emph
  {et~al.}(1998)\citenamefont {Mej\'ia-Monasterio}, \citenamefont {Richert},
  \citenamefont {Rupp},\ and\ \citenamefont {Weidenm\"uller}}]{Monasterio1998}%
  \BibitemOpen
  \bibfield  {author} {\bibinfo {author} {\bibfnamefont {C.}~\bibnamefont
  {Mej\'ia-Monasterio}}, \bibinfo {author} {\bibfnamefont {J.}~\bibnamefont
  {Richert}}, \bibinfo {author} {\bibfnamefont {T.}~\bibnamefont {Rupp}}, \
  and\ \bibinfo {author} {\bibfnamefont {H.~A.}\ \bibnamefont
  {Weidenm\"uller}},\ }\href {\doibase 10.1103/PhysRevLett.81.5189} {\bibfield
  {journal} {\bibinfo  {journal} {Phys. Rev. Lett.}\ }\textbf {\bibinfo
  {volume} {81}},\ \bibinfo {pages} {5189} (\bibinfo {year}
  {1998})}\BibitemShut {NoStop}%
\bibitem [{\citenamefont {{Oganesyan}}\ and\ \citenamefont
  {{Huse}}(2007)}]{Oganesyan07}%
  \BibitemOpen
  \bibfield  {author} {\bibinfo {author} {\bibfnamefont {V.}~\bibnamefont
  {{Oganesyan}}}\ and\ \bibinfo {author} {\bibfnamefont {D.~A.}\ \bibnamefont
  {{Huse}}},\ }\href {\doibase 10.1103/PhysRevB.75.155111} {\bibfield
  {journal} {\bibinfo  {journal} {\prb}\ }\textbf {\bibinfo {volume} {75}},\
  \bibinfo {pages} {155111} (\bibinfo {year} {2007})}\BibitemShut {NoStop}%
\bibitem [{\citenamefont {Pal}\ and\ \citenamefont {Huse}(2010)}]{Pal2010}%
  \BibitemOpen
  \bibfield  {author} {\bibinfo {author} {\bibfnamefont {A.}~\bibnamefont
  {Pal}}\ and\ \bibinfo {author} {\bibfnamefont {D.~A.}\ \bibnamefont {Huse}},\
  }\href {\doibase 10.1103/PhysRevB.82.174411} {\bibfield  {journal} {\bibinfo
  {journal} {Phys. Rev. B}\ }\textbf {\bibinfo {volume} {82}},\ \bibinfo
  {pages} {174411} (\bibinfo {year} {2010})}\BibitemShut {NoStop}%
\bibitem [{\citenamefont {{Iyer}}\ \emph {et~al.}(2012)\citenamefont {{Iyer}},
  \citenamefont {{Oganesyan}}, \citenamefont {{Refael}},\ and\ \citenamefont
  {{Huse}}}]{Iyer12}%
  \BibitemOpen
  \bibfield  {author} {\bibinfo {author} {\bibfnamefont {S.}~\bibnamefont
  {{Iyer}}}, \bibinfo {author} {\bibfnamefont {V.}~\bibnamefont {{Oganesyan}}},
  \bibinfo {author} {\bibfnamefont {G.}~\bibnamefont {{Refael}}}, \ and\
  \bibinfo {author} {\bibfnamefont {D.~A.}\ \bibnamefont {{Huse}}},\
  }\href@noop {} {\bibfield  {journal} {\bibinfo  {journal} {Preprint}\ }
  (\bibinfo {year} {2012})},\ \Eprint {http://arxiv.org/abs/1212.4159}
  {arXiv:1212.4159} \BibitemShut {NoStop}%
\bibitem [{\citenamefont {\u{Z}nidari\u{c}}\ \emph {et~al.}(2008)\citenamefont
  {\u{Z}nidari\u{c}}, \citenamefont {Prosen},\ and\ \citenamefont
  {Prelov\u{s}ek}}]{Znidaric08}%
  \BibitemOpen
  \bibfield  {author} {\bibinfo {author} {\bibfnamefont {M.}~\bibnamefont
  {\u{Z}nidari\u{c}}}, \bibinfo {author} {\bibfnamefont {T.}~\bibnamefont
  {Prosen}}, \ and\ \bibinfo {author} {\bibfnamefont {P.}~\bibnamefont
  {Prelov\u{s}ek}},\ }\href {\doibase 10.1103/PhysRevB.77.064426} {\bibfield
  {journal} {\bibinfo  {journal} {Phys. Rev. B}\ }\textbf {\bibinfo {volume}
  {77}},\ \bibinfo {pages} {064426} (\bibinfo {year} {2008})}\BibitemShut
  {NoStop}%
\bibitem [{\citenamefont {Chiara}\ \emph {et~al.}(2006)\citenamefont {Chiara},
  \citenamefont {Montangero}, \citenamefont {Calabrese},\ and\ \citenamefont
  {Fazio}}]{DeChiara2006}%
  \BibitemOpen
  \bibfield  {author} {\bibinfo {author} {\bibfnamefont {G.~D.}\ \bibnamefont
  {Chiara}}, \bibinfo {author} {\bibfnamefont {S.}~\bibnamefont {Montangero}},
  \bibinfo {author} {\bibfnamefont {P.}~\bibnamefont {Calabrese}}, \ and\
  \bibinfo {author} {\bibfnamefont {R.}~\bibnamefont {Fazio}},\ }\href
  {\doibase 10.1088/1742-5468/2006/03/P03001} {\bibfield  {journal} {\bibinfo
  {journal} {J. Stat. Mech.}\ ,\ \bibinfo {pages} {P03001}} (\bibinfo {year}
  {2006})}\BibitemShut {NoStop}%
\bibitem [{\citenamefont {{Bardarson}}\ \emph {et~al.}(2012)\citenamefont
  {{Bardarson}}, \citenamefont {{Pollmann}},\ and\ \citenamefont
  {{Moore}}}]{Bardason12}%
  \BibitemOpen
  \bibfield  {author} {\bibinfo {author} {\bibfnamefont {J.~H.}\ \bibnamefont
  {{Bardarson}}}, \bibinfo {author} {\bibfnamefont {F.}~\bibnamefont
  {{Pollmann}}}, \ and\ \bibinfo {author} {\bibfnamefont {J.~E.}\ \bibnamefont
  {{Moore}}},\ }\href {\doibase 10.1103/PhysRevLett.109.017202} {\bibfield
  {journal} {\bibinfo  {journal} {\prl}\ }\textbf {\bibinfo {volume} {109}},\
  \bibinfo {pages} {017202} (\bibinfo {year} {2012})}\BibitemShut {NoStop}%
\bibitem [{\citenamefont {{Vosk}}\ and\ \citenamefont
  {{Altman}}(2013)}]{Vosk13}%
  \BibitemOpen
  \bibfield  {author} {\bibinfo {author} {\bibfnamefont {R.}~\bibnamefont
  {{Vosk}}}\ and\ \bibinfo {author} {\bibfnamefont {E.}~\bibnamefont
  {{Altman}}},\ }\href {\doibase 10.1103/PhysRevLett.110.067204} {\bibfield
  {journal} {\bibinfo  {journal} {\prl}\ }\textbf {\bibinfo {volume} {110}},\
  \bibinfo {pages} {067204} (\bibinfo {year} {2013})}\BibitemShut {NoStop}%
\bibitem [{\citenamefont {Serbyn}\ \emph
  {et~al.}(2013{\natexlab{a}})\citenamefont {Serbyn}, \citenamefont
  {Papi\'{c}},\ and\ \citenamefont {Abanin}}]{serbyn2013-1}%
  \BibitemOpen
  \bibfield  {author} {\bibinfo {author} {\bibfnamefont {M.}~\bibnamefont
  {Serbyn}}, \bibinfo {author} {\bibfnamefont {Z.}~\bibnamefont {Papi\'{c}}}, \
  and\ \bibinfo {author} {\bibfnamefont {D.}~\bibnamefont {Abanin}},\
  }\href@noop {} {\bibfield  {journal} {\bibinfo  {journal} {Preprint (to
  appear in Phys. Rev. Lett.)}\ } (\bibinfo {year} {2013}{\natexlab{a}})},\
  \Eprint {http://arxiv.org/abs/1304.4605} {arXiv:1304.4605} \BibitemShut
  {NoStop}%
\bibitem [{\citenamefont {Huse}\ and\ \citenamefont
  {{Oganesyan}}(2013)}]{Huse2013}%
  \BibitemOpen
  \bibfield  {author} {\bibinfo {author} {\bibfnamefont {D.~A.}\ \bibnamefont
  {Huse}}\ and\ \bibinfo {author} {\bibfnamefont {V.}~\bibnamefont
  {{Oganesyan}}},\ }\href@noop {} {\bibfield  {journal} {\bibinfo  {journal}
  {Preprint}\ } (\bibinfo {year} {2013})},\ \Eprint
  {http://arxiv.org/abs/1305.4915} {arXiv:1305.4915} \BibitemShut {NoStop}%
\bibitem [{\citenamefont {Burrell}\ and\ \citenamefont
  {Osborne}(2007)}]{Burrell2007}%
  \BibitemOpen
  \bibfield  {author} {\bibinfo {author} {\bibfnamefont {C.~K.}\ \bibnamefont
  {Burrell}}\ and\ \bibinfo {author} {\bibfnamefont {T.~J.}\ \bibnamefont
  {Osborne}},\ }\href {\doibase 10.1103/PhysRevLett.99.167201} {\bibfield
  {journal} {\bibinfo  {journal} {Phys. Rev. Lett.}\ }\textbf {\bibinfo
  {volume} {99}},\ \bibinfo {pages} {167201} (\bibinfo {year}
  {2007})}\BibitemShut {NoStop}%
\bibitem [{\citenamefont {Fr\"ohlich}\ and\ \citenamefont
  {Spencer}(1983)}]{Frohlich83}%
  \BibitemOpen
  \bibfield  {author} {\bibinfo {author} {\bibfnamefont {J.}~\bibnamefont
  {Fr\"ohlich}}\ and\ \bibinfo {author} {\bibfnamefont {T.}~\bibnamefont
  {Spencer}},\ }\href {\doibase 10.1007/BF01209475} {\bibfield  {journal}
  {\bibinfo  {journal} {Comm. Math. Phys.}\ }\textbf {\bibinfo {volume} {88}},\
  \bibinfo {pages} {151} (\bibinfo {year} {1983})}\BibitemShut {NoStop}%
\bibitem [{\citenamefont {Eisert}\ \emph {et~al.}(2010)\citenamefont {Eisert},
  \citenamefont {Cramer},\ and\ \citenamefont {Plenio}}]{Eisert2010}%
  \BibitemOpen
  \bibfield  {author} {\bibinfo {author} {\bibfnamefont {J.}~\bibnamefont
  {Eisert}}, \bibinfo {author} {\bibfnamefont {M.}~\bibnamefont {Cramer}}, \
  and\ \bibinfo {author} {\bibfnamefont {M.~B.}\ \bibnamefont {Plenio}},\
  }\href {\doibase 10.1103/RevModPhys.82.277} {\bibfield  {journal} {\bibinfo
  {journal} {Rev. Mod. Phys.}\ }\textbf {\bibinfo {volume} {82}},\ \bibinfo
  {pages} {277} (\bibinfo {year} {2010})}\BibitemShut {NoStop}%
\bibitem [{\citenamefont {Berkovits}(2012)}]{Berkovits12}%
  \BibitemOpen
  \bibfield  {author} {\bibinfo {author} {\bibfnamefont {R.}~\bibnamefont
  {Berkovits}},\ }\href {\doibase 10.1103/PhysRevLett.108.176803} {\bibfield
  {journal} {\bibinfo  {journal} {Phys. Rev. Lett.}\ }\textbf {\bibinfo
  {volume} {108}},\ \bibinfo {pages} {176803} (\bibinfo {year}
  {2012})}\BibitemShut {NoStop}%
\bibitem [{\citenamefont {Wolf}\ \emph {et~al.}(2008)\citenamefont {Wolf},
  \citenamefont {Verstraete}, \citenamefont {Hastings},\ and\ \citenamefont
  {Cirac}}]{Wolf2008}%
  \BibitemOpen
  \bibfield  {author} {\bibinfo {author} {\bibfnamefont {M.~M.}\ \bibnamefont
  {Wolf}}, \bibinfo {author} {\bibfnamefont {F.}~\bibnamefont {Verstraete}},
  \bibinfo {author} {\bibfnamefont {M.~B.}\ \bibnamefont {Hastings}}, \ and\
  \bibinfo {author} {\bibfnamefont {J.~I.}\ \bibnamefont {Cirac}},\ }\href
  {\doibase 10.1103/PhysRevLett.100.070502} {\bibfield  {journal} {\bibinfo
  {journal} {Phys. Rev. Lett.}\ }\textbf {\bibinfo {volume} {100}},\ \bibinfo
  {pages} {070502} (\bibinfo {year} {2008})}\BibitemShut {NoStop}%
\bibitem [{\citenamefont {Vidal}\ \emph {et~al.}(2003)\citenamefont {Vidal},
  \citenamefont {Latorre}, \citenamefont {Rico},\ and\ \citenamefont
  {Kitaev}}]{Vidal2003}%
  \BibitemOpen
  \bibfield  {author} {\bibinfo {author} {\bibfnamefont {G.}~\bibnamefont
  {Vidal}}, \bibinfo {author} {\bibfnamefont {J.~I.}\ \bibnamefont {Latorre}},
  \bibinfo {author} {\bibfnamefont {E.}~\bibnamefont {Rico}}, \ and\ \bibinfo
  {author} {\bibfnamefont {A.}~\bibnamefont {Kitaev}},\ }\href {\doibase
  10.1103/PhysRevLett.90.227902} {\bibfield  {journal} {\bibinfo  {journal}
  {Phys. Rev. Lett.}\ }\textbf {\bibinfo {volume} {90}},\ \bibinfo {pages}
  {227902} (\bibinfo {year} {2003})}\BibitemShut {NoStop}%
\bibitem [{\citenamefont {{Chakravarty}}(2010)}]{Chakravarty10}%
  \BibitemOpen
  \bibfield  {author} {\bibinfo {author} {\bibfnamefont {S.}~\bibnamefont
  {{Chakravarty}}},\ }\href {\doibase 10.1142/S0217979210064629} {\bibfield
  {journal} {\bibinfo  {journal} {Int. J. Mod. Phys. B}\ }\textbf {\bibinfo
  {volume} {24}},\ \bibinfo {pages} {1823} (\bibinfo {year}
  {2010})}\BibitemShut {NoStop}%
\bibitem [{\citenamefont {{Huse}}\ \emph {et~al.}(2013)\citenamefont {{Huse}},
  \citenamefont {{Nandkishore}}, \citenamefont {{Oganesyan}}, \citenamefont
  {{Pal}},\ and\ \citenamefont {{Sondhi}}}]{Huse13}%
  \BibitemOpen
  \bibfield  {author} {\bibinfo {author} {\bibfnamefont {D.~A.}\ \bibnamefont
  {{Huse}}}, \bibinfo {author} {\bibfnamefont {R.}~\bibnamefont
  {{Nandkishore}}}, \bibinfo {author} {\bibfnamefont {V.}~\bibnamefont
  {{Oganesyan}}}, \bibinfo {author} {\bibfnamefont {A.}~\bibnamefont {{Pal}}},
  \ and\ \bibinfo {author} {\bibfnamefont {S.~L.}\ \bibnamefont {{Sondhi}}},\
  }\href@noop {} {\bibfield  {journal} {\bibinfo  {journal} {Preprint}\ }
  (\bibinfo {year} {2013})},\ \Eprint {http://arxiv.org/abs/1304.1158}
  {arXiv:1304.1158} \BibitemShut {NoStop}%
\bibitem [{\citenamefont {Monthus}\ and\ \citenamefont
  {Garel}(2010)}]{Monthus2010}%
  \BibitemOpen
  \bibfield  {author} {\bibinfo {author} {\bibfnamefont {C.}~\bibnamefont
  {Monthus}}\ and\ \bibinfo {author} {\bibfnamefont {T.}~\bibnamefont
  {Garel}},\ }\href {\doibase 10.1103/PhysRevB.81.134202} {\bibfield  {journal}
  {\bibinfo  {journal} {Phys. Rev. B}\ }\textbf {\bibinfo {volume} {81}},\
  \bibinfo {pages} {134202} (\bibinfo {year} {2010})}\BibitemShut {NoStop}%
\bibitem [{\citenamefont {Aleiner}\ \emph {et~al.}(2002)\citenamefont
  {Aleiner}, \citenamefont {Brouwer},\ and\ \citenamefont
  {Glazman}}]{Aleiner02}%
  \BibitemOpen
  \bibfield  {author} {\bibinfo {author} {\bibfnamefont {I.}~\bibnamefont
  {Aleiner}}, \bibinfo {author} {\bibfnamefont {P.}~\bibnamefont {Brouwer}}, \
  and\ \bibinfo {author} {\bibfnamefont {L.}~\bibnamefont {Glazman}},\ }\href
  {\doibase 10.1016/S0370-1573(01)00063-1} {\bibfield  {journal} {\bibinfo
  {journal} {{Physics Reports}}\ }\textbf {\bibinfo {volume} {358}},\ \bibinfo
  {pages} {309} (\bibinfo {year} {2002})}\BibitemShut {NoStop}%
\bibitem [{\citenamefont {Mirlin}(2000)}]{Mirlin00}%
  \BibitemOpen
  \bibfield  {author} {\bibinfo {author} {\bibfnamefont {A.~D.}\ \bibnamefont
  {Mirlin}},\ }\href {\doibase 10.1016/S0370-1573(99)00091-5} {\bibfield
  {journal} {\bibinfo  {journal} {{Physics Reports}}\ }\textbf {\bibinfo
  {volume} {326}},\ \bibinfo {pages} {259} (\bibinfo {year}
  {2000})}\BibitemShut {NoStop}%
\bibitem [{\citenamefont {Helstrom}(1976)}]{Helstrom1976}%
  \BibitemOpen
  \bibfield  {author} {\bibinfo {author} {\bibfnamefont {C.}~\bibnamefont
  {Helstrom}},\ }\enquote {\bibinfo {title} {Quantum detection and estimation
  theory},}\ \ (\bibinfo  {publisher} {Academic},\ \bibinfo {year}
  {1976})\BibitemShut {NoStop}%
\bibitem [{\citenamefont {Fuchs}\ and\ \citenamefont {van~de
  Graaf}(1999)}]{Fuchs1999}%
  \BibitemOpen
  \bibfield  {author} {\bibinfo {author} {\bibfnamefont {C.}~\bibnamefont
  {Fuchs}}\ and\ \bibinfo {author} {\bibfnamefont {J.}~\bibnamefont {van~de
  Graaf}},\ }\href {\doibase 10.1109/18.761271} {\bibfield  {journal} {\bibinfo
   {journal} {IEEE\ Trans.\ Inf.\ Theory}\ }\textbf {\bibinfo {volume} {45}},\
  \bibinfo {pages} {1216} (\bibinfo {year} {1999})}\BibitemShut {NoStop}%
\bibitem [{\citenamefont {Bauer}\ \emph {et~al.}(2010)\citenamefont {Bauer},
  \citenamefont {Troyer}, \citenamefont {Scarola},\ and\ \citenamefont
  {Whaley}}]{Bauer2010}%
  \BibitemOpen
  \bibfield  {author} {\bibinfo {author} {\bibfnamefont {B.}~\bibnamefont
  {Bauer}}, \bibinfo {author} {\bibfnamefont {M.}~\bibnamefont {Troyer}},
  \bibinfo {author} {\bibfnamefont {V.~W.}\ \bibnamefont {Scarola}}, \ and\
  \bibinfo {author} {\bibfnamefont {K.~B.}\ \bibnamefont {Whaley}},\ }\href
  {\doibase 10.1103/PhysRevB.81.085118} {\bibfield  {journal} {\bibinfo
  {journal} {Phys. Rev. B}\ }\textbf {\bibinfo {volume} {81}},\ \bibinfo
  {pages} {085118} (\bibinfo {year} {2010})}\BibitemShut {NoStop}%
\bibitem [{\citenamefont {Verstraete}\ and\ \citenamefont
  {Cirac}(2006)}]{Verstraete2006}%
  \BibitemOpen
  \bibfield  {author} {\bibinfo {author} {\bibfnamefont {F.}~\bibnamefont
  {Verstraete}}\ and\ \bibinfo {author} {\bibfnamefont {J.~I.}\ \bibnamefont
  {Cirac}},\ }\href {\doibase 10.1103/PhysRevB.73.094423} {\bibfield  {journal}
  {\bibinfo  {journal} {Phys. Rev. B}\ }\textbf {\bibinfo {volume} {73}},\
  \bibinfo {pages} {094423} (\bibinfo {year} {2006})}\BibitemShut {NoStop}%
\bibitem [{\citenamefont {Hastings}(2007{\natexlab{a}})}]{Hastings2007-prb}%
  \BibitemOpen
  \bibfield  {author} {\bibinfo {author} {\bibfnamefont {M.~B.}\ \bibnamefont
  {Hastings}},\ }\href {\doibase 10.1103/PhysRevB.76.035114} {\bibfield
  {journal} {\bibinfo  {journal} {Phys. Rev. B}\ }\textbf {\bibinfo {volume}
  {76}},\ \bibinfo {pages} {035114} (\bibinfo {year}
  {2007}{\natexlab{a}})}\BibitemShut {NoStop}%
\bibitem [{\citenamefont {Hastings}(2007{\natexlab{b}})}]{Hastings2007}%
  \BibitemOpen
  \bibfield  {author} {\bibinfo {author} {\bibfnamefont {M.~B.}\ \bibnamefont
  {Hastings}},\ }\href {\doibase 10.1088/1742-5468/2007/08/P08024} {\bibfield
  {journal} {\bibinfo  {journal} {J. Stat. Mech.}\ ,\ \bibinfo {pages}
  {P08024}} (\bibinfo {year} {2007}{\natexlab{b}})}\BibitemShut {NoStop}%
\bibitem [{\citenamefont {Serbyn}\ \emph
  {et~al.}(2013{\natexlab{b}})\citenamefont {Serbyn}, \citenamefont
  {Papi\'{c}},\ and\ \citenamefont {Abanin}}]{Serbyn2013}%
  \BibitemOpen
  \bibfield  {author} {\bibinfo {author} {\bibfnamefont {M.}~\bibnamefont
  {Serbyn}}, \bibinfo {author} {\bibfnamefont {Z.}~\bibnamefont {Papi\'{c}}}, \
  and\ \bibinfo {author} {\bibfnamefont {D.}~\bibnamefont {Abanin}},\
  }\href@noop {} {\bibfield  {journal} {\bibinfo  {journal} {Preprint}\ }
  (\bibinfo {year} {2013}{\natexlab{b}})},\ \Eprint
  {http://arxiv.org/abs/1305.5554} {arXiv:1305.5554} \BibitemShut {NoStop}%
\bibitem [{\citenamefont {Gu}\ and\ \citenamefont {Wen}(2009)}]{gu2009}%
  \BibitemOpen
  \bibfield  {author} {\bibinfo {author} {\bibfnamefont {Z.-C.}\ \bibnamefont
  {Gu}}\ and\ \bibinfo {author} {\bibfnamefont {X.-G.}\ \bibnamefont {Wen}},\
  }\href {\doibase 10.1103/PhysRevB.80.155131} {\bibfield  {journal} {\bibinfo
  {journal} {Phys. Rev. B}\ }\textbf {\bibinfo {volume} {80}},\ \bibinfo
  {pages} {155131} (\bibinfo {year} {2009})}\BibitemShut {NoStop}%
\bibitem [{Note1()}]{Note1}%
  \BibitemOpen
  \bibinfo {note} {Note that for an SPT phase, such a decomposition is possible
  if one first applies a finite-depth unitary transformation that breaks the
  symmetry.}\BibitemShut {Stop}%
\bibitem [{\citenamefont {{Rigol}}\ and\ \citenamefont
  {{Srednicki}}(2012)}]{Rigol12}%
  \BibitemOpen
  \bibfield  {author} {\bibinfo {author} {\bibfnamefont {M.}~\bibnamefont
  {{Rigol}}}\ and\ \bibinfo {author} {\bibfnamefont {M.}~\bibnamefont
  {{Srednicki}}},\ }\href {\doibase 10.1103/PhysRevLett.108.110601} {\bibfield
  {journal} {\bibinfo  {journal} {Phys. Rev. Lett.}\ }\textbf {\bibinfo
  {volume} {108}},\ \bibinfo {pages} {110601} (\bibinfo {year}
  {2012})}\BibitemShut {NoStop}%
\bibitem [{\citenamefont {Bekenstein}(1973)}]{Bekenstein1973}%
  \BibitemOpen
  \bibfield  {author} {\bibinfo {author} {\bibfnamefont {J.~D.}\ \bibnamefont
  {Bekenstein}},\ }\href {\doibase 10.1103/PhysRevD.7.2333} {\bibfield
  {journal} {\bibinfo  {journal} {Phys. Rev. D}\ }\textbf {\bibinfo {volume}
  {7}},\ \bibinfo {pages} {2333} (\bibinfo {year} {1973})}\BibitemShut
  {NoStop}%
\bibitem [{\citenamefont {Bombelli}\ \emph {et~al.}(1986)\citenamefont
  {Bombelli}, \citenamefont {Koul}, \citenamefont {Lee},\ and\ \citenamefont
  {Sorkin}}]{Bombelli1986}%
  \BibitemOpen
  \bibfield  {author} {\bibinfo {author} {\bibfnamefont {L.}~\bibnamefont
  {Bombelli}}, \bibinfo {author} {\bibfnamefont {R.~K.}\ \bibnamefont {Koul}},
  \bibinfo {author} {\bibfnamefont {J.}~\bibnamefont {Lee}}, \ and\ \bibinfo
  {author} {\bibfnamefont {R.~D.}\ \bibnamefont {Sorkin}},\ }\href {\doibase
  10.1103/PhysRevD.34.373} {\bibfield  {journal} {\bibinfo  {journal} {Phys.
  Rev. D}\ }\textbf {\bibinfo {volume} {34}},\ \bibinfo {pages} {373} (\bibinfo
  {year} {1986})}\BibitemShut {NoStop}%
\bibitem [{\citenamefont {Srednicki}(1993)}]{Srednicki1993}%
  \BibitemOpen
  \bibfield  {author} {\bibinfo {author} {\bibfnamefont {M.}~\bibnamefont
  {Srednicki}},\ }\href {\doibase 10.1103/PhysRevLett.71.666} {\bibfield
  {journal} {\bibinfo  {journal} {Phys. Rev. Lett.}\ }\textbf {\bibinfo
  {volume} {71}},\ \bibinfo {pages} {666} (\bibinfo {year} {1993})}\BibitemShut
  {NoStop}%
\bibitem [{\citenamefont {Holzhey}\ \emph {et~al.}(1994)\citenamefont
  {Holzhey}, \citenamefont {Larsen},\ and\ \citenamefont
  {Wilczek}}]{holzhey1994}%
  \BibitemOpen
  \bibfield  {author} {\bibinfo {author} {\bibfnamefont {C.}~\bibnamefont
  {Holzhey}}, \bibinfo {author} {\bibfnamefont {F.}~\bibnamefont {Larsen}}, \
  and\ \bibinfo {author} {\bibfnamefont {F.}~\bibnamefont {Wilczek}},\ }\href
  {\doibase DOI: 10.1016/0550-3213(94)90402-2} {\bibfield  {journal} {\bibinfo
  {journal} {Nucl. Phys. B}\ }\textbf {\bibinfo {volume} {424}},\ \bibinfo
  {pages} {443 } (\bibinfo {year} {1994})}\BibitemShut {NoStop}%
\bibitem [{\citenamefont {Calabrese}\ and\ \citenamefont
  {Cardy}(2004)}]{calabrese2004}%
  \BibitemOpen
  \bibfield  {author} {\bibinfo {author} {\bibfnamefont {P.}~\bibnamefont
  {Calabrese}}\ and\ \bibinfo {author} {\bibfnamefont {J.}~\bibnamefont
  {Cardy}},\ }\href {\doibase 10.1088/1742-5468/2004/06/P06002} {\bibfield
  {journal} {\bibinfo  {journal} {J. Stat. Mech.}\ ,\ \bibinfo {pages}
  {P06002}} (\bibinfo {year} {2004})}\BibitemShut {NoStop}%
\bibitem [{\citenamefont {Refael}\ and\ \citenamefont
  {Moore}(2004)}]{Refael04}%
  \BibitemOpen
  \bibfield  {author} {\bibinfo {author} {\bibfnamefont {G.}~\bibnamefont
  {Refael}}\ and\ \bibinfo {author} {\bibfnamefont {J.~E.}\ \bibnamefont
  {Moore}},\ }\href {\doibase 10.1103/PhysRevLett.93.260602} {\bibfield
  {journal} {\bibinfo  {journal} {Phys. Rev. Lett.}\ }\textbf {\bibinfo
  {volume} {93}},\ \bibinfo {pages} {260602} (\bibinfo {year}
  {2004})}\BibitemShut {NoStop}%
\bibitem [{\citenamefont {Plenio}\ \emph {et~al.}(2005)\citenamefont {Plenio},
  \citenamefont {Eisert}, \citenamefont {Drei\ss{}ig},\ and\ \citenamefont
  {Cramer}}]{Plenio2005}%
  \BibitemOpen
  \bibfield  {author} {\bibinfo {author} {\bibfnamefont {M.~B.}\ \bibnamefont
  {Plenio}}, \bibinfo {author} {\bibfnamefont {J.}~\bibnamefont {Eisert}},
  \bibinfo {author} {\bibfnamefont {J.}~\bibnamefont {Drei\ss{}ig}}, \ and\
  \bibinfo {author} {\bibfnamefont {M.}~\bibnamefont {Cramer}},\ }\href
  {\doibase 10.1103/PhysRevLett.94.060503} {\bibfield  {journal} {\bibinfo
  {journal} {Phys. Rev. Lett.}\ }\textbf {\bibinfo {volume} {94}},\ \bibinfo
  {pages} {060503} (\bibinfo {year} {2005})}\BibitemShut {NoStop}%
\bibitem [{\citenamefont {Wolf}(2006)}]{Wolf06}%
  \BibitemOpen
  \bibfield  {author} {\bibinfo {author} {\bibfnamefont {M.~M.}\ \bibnamefont
  {Wolf}},\ }\href {\doibase 10.1103/PhysRevLett.96.010404} {\bibfield
  {journal} {\bibinfo  {journal} {Phys. Rev. Lett.}\ }\textbf {\bibinfo
  {volume} {96}},\ \bibinfo {pages} {010404} (\bibinfo {year}
  {2006})}\BibitemShut {NoStop}%
\bibitem [{\citenamefont {Gioev}\ and\ \citenamefont {Klich}(2006)}]{Gioev06}%
  \BibitemOpen
  \bibfield  {author} {\bibinfo {author} {\bibfnamefont {D.}~\bibnamefont
  {Gioev}}\ and\ \bibinfo {author} {\bibfnamefont {I.}~\bibnamefont {Klich}},\
  }\href {\doibase 10.1103/PhysRevLett.96.100503} {\bibfield  {journal}
  {\bibinfo  {journal} {Phys. Rev. Lett.}\ }\textbf {\bibinfo {volume} {96}},\
  \bibinfo {pages} {100503} (\bibinfo {year} {2006})}\BibitemShut {NoStop}%
\bibitem [{\citenamefont {{Hastings}}(2011)}]{Hastings11}%
  \BibitemOpen
  \bibfield  {author} {\bibinfo {author} {\bibfnamefont {M.~B.}\ \bibnamefont
  {{Hastings}}},\ }\href {\doibase 10.1103/PhysRevLett.107.210501} {\bibfield
  {journal} {\bibinfo  {journal} {Phys. Rev. Lett.}\ }\textbf {\bibinfo
  {volume} {107}},\ \bibinfo {pages} {210501} (\bibinfo {year}
  {2011})}\BibitemShut {NoStop}%
\bibitem [{\citenamefont {Hastings}\ and\ \citenamefont
  {Wen}(2005)}]{Hastings05}%
  \BibitemOpen
  \bibfield  {author} {\bibinfo {author} {\bibfnamefont {M.~B.}\ \bibnamefont
  {Hastings}}\ and\ \bibinfo {author} {\bibfnamefont {X.-G.}\ \bibnamefont
  {Wen}},\ }\href {\doibase 10.1103/PhysRevB.72.045141} {\bibfield  {journal}
  {\bibinfo  {journal} {Phys. Rev. B}\ }\textbf {\bibinfo {volume} {72}},\
  \bibinfo {pages} {045141} (\bibinfo {year} {2005})}\BibitemShut {NoStop}%
\bibitem [{\citenamefont {Nayak}\ \emph {et~al.}(2008)\citenamefont {Nayak},
  \citenamefont {Simon}, \citenamefont {Stern}, \citenamefont {Freedman},\ and\
  \citenamefont {Sarma}}]{Nayak08}%
  \BibitemOpen
  \bibfield  {author} {\bibinfo {author} {\bibfnamefont {C.}~\bibnamefont
  {Nayak}}, \bibinfo {author} {\bibfnamefont {S.~H.}\ \bibnamefont {Simon}},
  \bibinfo {author} {\bibfnamefont {A.}~\bibnamefont {Stern}}, \bibinfo
  {author} {\bibfnamefont {M.}~\bibnamefont {Freedman}}, \ and\ \bibinfo
  {author} {\bibfnamefont {S.~D.}\ \bibnamefont {Sarma}},\ }\href {\doibase
  10.1103/RevModPhys.80.1083} {\bibfield  {journal} {\bibinfo  {journal} {Rev.
  Mod. Phys.}\ }\textbf {\bibinfo {volume} {80}},\ \bibinfo {pages} {1083}
  (\bibinfo {year} {2008})}\BibitemShut {NoStop}%
\bibitem [{\citenamefont {Kitaev}(2006)}]{Kitaev2006}%
  \BibitemOpen
  \bibfield  {author} {\bibinfo {author} {\bibfnamefont {A.}~\bibnamefont
  {Kitaev}},\ }\href {\doibase 10.1016/j.aop.2005.10.005} {\bibfield  {journal}
  {\bibinfo  {journal} {Annals of Physics}\ }\textbf {\bibinfo {volume}
  {321}},\ \bibinfo {pages} {2 } (\bibinfo {year} {2006})}\BibitemShut
  {NoStop}%
\bibitem [{\citenamefont {Trebst}\ \emph {et~al.}(2007)\citenamefont {Trebst},
  \citenamefont {Werner}, \citenamefont {Troyer}, \citenamefont {Shtengel},\
  and\ \citenamefont {Nayak}}]{Trebst2007}%
  \BibitemOpen
  \bibfield  {author} {\bibinfo {author} {\bibfnamefont {S.}~\bibnamefont
  {Trebst}}, \bibinfo {author} {\bibfnamefont {P.}~\bibnamefont {Werner}},
  \bibinfo {author} {\bibfnamefont {M.}~\bibnamefont {Troyer}}, \bibinfo
  {author} {\bibfnamefont {K.}~\bibnamefont {Shtengel}}, \ and\ \bibinfo
  {author} {\bibfnamefont {C.}~\bibnamefont {Nayak}},\ }\href {\doibase
  10.1103/PhysRevLett.98.070602} {\bibfield  {journal} {\bibinfo  {journal}
  {Phys. Rev. Lett.}\ }\textbf {\bibinfo {volume} {98}},\ \bibinfo {pages}
  {070602} (\bibinfo {year} {2007})}\BibitemShut {NoStop}%
\bibitem [{\citenamefont {Hamma}\ and\ \citenamefont
  {Lidar}(2008)}]{Hamma2008}%
  \BibitemOpen
  \bibfield  {author} {\bibinfo {author} {\bibfnamefont {A.}~\bibnamefont
  {Hamma}}\ and\ \bibinfo {author} {\bibfnamefont {D.~A.}\ \bibnamefont
  {Lidar}},\ }\href {\doibase 10.1103/PhysRevLett.100.030502} {\bibfield
  {journal} {\bibinfo  {journal} {Phys. Rev. Lett.}\ }\textbf {\bibinfo
  {volume} {100}},\ \bibinfo {pages} {030502} (\bibinfo {year}
  {2008})}\BibitemShut {NoStop}%
\bibitem [{\citenamefont {Vidal}\ \emph
  {et~al.}(2009{\natexlab{a}})\citenamefont {Vidal}, \citenamefont {Dusuel},\
  and\ \citenamefont {Schmidt}}]{Vidal2009}%
  \BibitemOpen
  \bibfield  {author} {\bibinfo {author} {\bibfnamefont {J.}~\bibnamefont
  {Vidal}}, \bibinfo {author} {\bibfnamefont {S.}~\bibnamefont {Dusuel}}, \
  and\ \bibinfo {author} {\bibfnamefont {K.~P.}\ \bibnamefont {Schmidt}},\
  }\href {\doibase 10.1103/PhysRevB.79.033109} {\bibfield  {journal} {\bibinfo
  {journal} {Phys. Rev. B}\ }\textbf {\bibinfo {volume} {79}},\ \bibinfo
  {pages} {033109} (\bibinfo {year} {2009}{\natexlab{a}})}\BibitemShut
  {NoStop}%
\bibitem [{\citenamefont {Vidal}\ \emph
  {et~al.}(2009{\natexlab{b}})\citenamefont {Vidal}, \citenamefont {Thomale},
  \citenamefont {Schmidt},\ and\ \citenamefont {Dusuel}}]{Vidal2009-1}%
  \BibitemOpen
  \bibfield  {author} {\bibinfo {author} {\bibfnamefont {J.}~\bibnamefont
  {Vidal}}, \bibinfo {author} {\bibfnamefont {R.}~\bibnamefont {Thomale}},
  \bibinfo {author} {\bibfnamefont {K.~P.}\ \bibnamefont {Schmidt}}, \ and\
  \bibinfo {author} {\bibfnamefont {S.}~\bibnamefont {Dusuel}},\ }\href
  {\doibase 10.1103/PhysRevB.80.081104} {\bibfield  {journal} {\bibinfo
  {journal} {Phys. Rev. B}\ }\textbf {\bibinfo {volume} {80}},\ \bibinfo
  {pages} {081104} (\bibinfo {year} {2009}{\natexlab{b}})}\BibitemShut
  {NoStop}%
\bibitem [{\citenamefont {Tupitsyn}\ \emph {et~al.}(2010)\citenamefont
  {Tupitsyn}, \citenamefont {Kitaev}, \citenamefont {Prokof'ev},\ and\
  \citenamefont {Stamp}}]{Tupitsyn2010}%
  \BibitemOpen
  \bibfield  {author} {\bibinfo {author} {\bibfnamefont {I.~S.}\ \bibnamefont
  {Tupitsyn}}, \bibinfo {author} {\bibfnamefont {A.}~\bibnamefont {Kitaev}},
  \bibinfo {author} {\bibfnamefont {N.~V.}\ \bibnamefont {Prokof'ev}}, \ and\
  \bibinfo {author} {\bibfnamefont {P.~C.~E.}\ \bibnamefont {Stamp}},\ }\href
  {\doibase 10.1103/PhysRevB.82.085114} {\bibfield  {journal} {\bibinfo
  {journal} {Phys. Rev. B}\ }\textbf {\bibinfo {volume} {82}},\ \bibinfo
  {pages} {085114} (\bibinfo {year} {2010})}\BibitemShut {NoStop}%
\bibitem [{\citenamefont {Dusuel}\ \emph {et~al.}(2011)\citenamefont {Dusuel},
  \citenamefont {Kamfor}, \citenamefont {Or\'us}, \citenamefont {Schmidt},\
  and\ \citenamefont {Vidal}}]{Dusuel2011}%
  \BibitemOpen
  \bibfield  {author} {\bibinfo {author} {\bibfnamefont {S.}~\bibnamefont
  {Dusuel}}, \bibinfo {author} {\bibfnamefont {M.}~\bibnamefont {Kamfor}},
  \bibinfo {author} {\bibfnamefont {R.}~\bibnamefont {Or\'us}}, \bibinfo
  {author} {\bibfnamefont {K.~P.}\ \bibnamefont {Schmidt}}, \ and\ \bibinfo
  {author} {\bibfnamefont {J.}~\bibnamefont {Vidal}},\ }\href {\doibase
  10.1103/PhysRevLett.106.107203} {\bibfield  {journal} {\bibinfo  {journal}
  {Phys. Rev. Lett.}\ }\textbf {\bibinfo {volume} {106}},\ \bibinfo {pages}
  {107203} (\bibinfo {year} {2011})}\BibitemShut {NoStop}%
\bibitem [{\citenamefont {Stark}\ \emph {et~al.}(2011)\citenamefont {Stark},
  \citenamefont {Pollet}, \citenamefont {Imamo\u{g}lu},\ and\ \citenamefont
  {Renner}}]{Stark11}%
  \BibitemOpen
  \bibfield  {author} {\bibinfo {author} {\bibfnamefont {C.}~\bibnamefont
  {Stark}}, \bibinfo {author} {\bibfnamefont {L.}~\bibnamefont {Pollet}},
  \bibinfo {author} {\bibfnamefont {A.}~\bibnamefont {Imamo\u{g}lu}}, \ and\
  \bibinfo {author} {\bibfnamefont {R.}~\bibnamefont {Renner}},\ }\href
  {\doibase 10.1103/PhysRevLett.107.030504} {\bibfield  {journal} {\bibinfo
  {journal} {Phys. Rev. Lett.}\ }\textbf {\bibinfo {volume} {107}},\ \bibinfo
  {pages} {030504} (\bibinfo {year} {2011})}\BibitemShut {NoStop}%
\bibitem [{\citenamefont {Bauer}\ and\ \citenamefont {Nayak}(2013)}]{Bauer13}%
  \BibitemOpen
  \bibfield  {author} {\bibinfo {author} {\bibfnamefont {B.}~\bibnamefont
  {Bauer}}\ and\ \bibinfo {author} {\bibfnamefont {C.}~\bibnamefont {Nayak}},\
  }\href@noop {} {\  (\bibinfo {year} {2013})}\BibitemShut {NoStop}%
\bibitem [{\citenamefont {Wootton}\ and\ \citenamefont
  {Pachos}(2011)}]{Wootton2011}%
  \BibitemOpen
  \bibfield  {author} {\bibinfo {author} {\bibfnamefont {J.~R.}\ \bibnamefont
  {Wootton}}\ and\ \bibinfo {author} {\bibfnamefont {J.~K.}\ \bibnamefont
  {Pachos}},\ }\href {\doibase 10.1103/PhysRevLett.107.030503} {\bibfield
  {journal} {\bibinfo  {journal} {Phys. Rev. Lett.}\ }\textbf {\bibinfo
  {volume} {107}},\ \bibinfo {pages} {030503} (\bibinfo {year}
  {2011})}\BibitemShut {NoStop}%
\bibitem [{\citenamefont {Bravyi}\ and\ \citenamefont
  {K\"onig}(2012)}]{Bravyi12}%
  \BibitemOpen
  \bibfield  {author} {\bibinfo {author} {\bibfnamefont {S.}~\bibnamefont
  {Bravyi}}\ and\ \bibinfo {author} {\bibfnamefont {R.}~\bibnamefont
  {K\"onig}},\ }\href {\doibase 10.1007/s00220-012-1606-9} {\bibfield
  {journal} {\bibinfo  {journal} {Comm. Math. Phys.}\ }\textbf {\bibinfo
  {volume} {316}},\ \bibinfo {pages} {641} (\bibinfo {year}
  {2012})}\BibitemShut {NoStop}%
\bibitem [{\citenamefont {Dennis}\ \emph {et~al.}(2002)\citenamefont {Dennis},
  \citenamefont {Kitaev}, \citenamefont {Landahl},\ and\ \citenamefont
  {Preskill}}]{Dennis2002}%
  \BibitemOpen
  \bibfield  {author} {\bibinfo {author} {\bibfnamefont {E.}~\bibnamefont
  {Dennis}}, \bibinfo {author} {\bibfnamefont {A.}~\bibnamefont {Kitaev}},
  \bibinfo {author} {\bibfnamefont {A.}~\bibnamefont {Landahl}}, \ and\
  \bibinfo {author} {\bibfnamefont {J.}~\bibnamefont {Preskill}},\ }\href
  {\doibase 10.1063/1.1499754} {\bibfield  {journal} {\bibinfo  {journal} {J.
  Math. Phys.}\ }\textbf {\bibinfo {volume} {43}},\ \bibinfo {pages} {4452}
  (\bibinfo {year} {2002})}\BibitemShut {NoStop}%
\bibitem [{\citenamefont {Alicki}\ \emph {et~al.}(2010)\citenamefont {Alicki},
  \citenamefont {Horodecki}, \citenamefont {Horodecki},\ and\ \citenamefont
  {Horodecki}}]{Alicki2010}%
  \BibitemOpen
  \bibfield  {author} {\bibinfo {author} {\bibfnamefont {R.}~\bibnamefont
  {Alicki}}, \bibinfo {author} {\bibfnamefont {M.}~\bibnamefont {Horodecki}},
  \bibinfo {author} {\bibfnamefont {P.}~\bibnamefont {Horodecki}}, \ and\
  \bibinfo {author} {\bibfnamefont {R.}~\bibnamefont {Horodecki}},\ }\href
  {\doibase 10.1142/S1230161210000023} {\bibfield  {journal} {\bibinfo
  {journal} {Open Syst. Inf. Dyn.}\ }\textbf {\bibinfo {volume} {17}},\
  \bibinfo {pages} {1} (\bibinfo {year} {2010})}\BibitemShut {NoStop}%
\bibitem [{\citenamefont {Alicki}\ \emph {et~al.}(2008)\citenamefont {Alicki},
  \citenamefont {Fannes},\ and\ \citenamefont {Horodecki}}]{Alicki2008}%
  \BibitemOpen
  \bibfield  {author} {\bibinfo {author} {\bibfnamefont {R.}~\bibnamefont
  {Alicki}}, \bibinfo {author} {\bibfnamefont {M.}~\bibnamefont {Fannes}}, \
  and\ \bibinfo {author} {\bibfnamefont {M.}~\bibnamefont {Horodecki}},\ }\href
  {\doibase 10.1088/1751-8113/42/6/065303} {\bibfield  {journal} {\bibinfo
  {journal} {J. Phys. A: Math. Theor.}\ }\textbf {\bibinfo {volume} {42}},\
  \bibinfo {pages} {065303} (\bibinfo {year} {2008})}\BibitemShut {NoStop}%
\bibitem [{\citenamefont {Alicki}\ and\ \citenamefont
  {Horodecki}(2006)}]{Alicki2006}%
  \BibitemOpen
  \bibfield  {author} {\bibinfo {author} {\bibfnamefont {R.}~\bibnamefont
  {Alicki}}\ and\ \bibinfo {author} {\bibfnamefont {M.}~\bibnamefont
  {Horodecki}},\ }\href@noop {} {\bibfield  {journal} {\bibinfo  {journal}
  {Preprint}\ } (\bibinfo {year} {2006})},\ \Eprint
  {http://arxiv.org/abs/quant-ph/0603260} {arXiv:quant-ph/0603260} \BibitemShut
  {NoStop}%
\bibitem [{\citenamefont {Bravyi}\ \emph {et~al.}(2006)\citenamefont {Bravyi},
  \citenamefont {Hastings},\ and\ \citenamefont {Verstraete}}]{Bravyi2006}%
  \BibitemOpen
  \bibfield  {author} {\bibinfo {author} {\bibfnamefont {S.}~\bibnamefont
  {Bravyi}}, \bibinfo {author} {\bibfnamefont {M.~B.}\ \bibnamefont
  {Hastings}}, \ and\ \bibinfo {author} {\bibfnamefont {F.}~\bibnamefont
  {Verstraete}},\ }\href {\doibase 10.1103/PhysRevLett.97.050401} {\bibfield
  {journal} {\bibinfo  {journal} {Phys. Rev. Lett.}\ }\textbf {\bibinfo
  {volume} {97}},\ \bibinfo {pages} {050401} (\bibinfo {year}
  {2006})}\BibitemShut {NoStop}%
\bibitem [{\citenamefont {Levin}\ and\ \citenamefont {Wen}(2006)}]{Levin06}%
  \BibitemOpen
  \bibfield  {author} {\bibinfo {author} {\bibfnamefont {M.}~\bibnamefont
  {Levin}}\ and\ \bibinfo {author} {\bibfnamefont {X.-G.}\ \bibnamefont
  {Wen}},\ }\href {\doibase 10.1103/PhysRevLett.96.110405} {\bibfield
  {journal} {\bibinfo  {journal} {Phys. Rev. Lett.}\ }\textbf {\bibinfo
  {volume} {96}},\ \bibinfo {pages} {110405} (\bibinfo {year}
  {2006})}\BibitemShut {NoStop}%
\bibitem [{\citenamefont {Bauer}\ \emph {et~al.}(2011)\citenamefont {Bauer}
  \emph {et~al.}}]{bauer2011-alps}%
  \BibitemOpen
  \bibfield  {author} {\bibinfo {author} {\bibfnamefont {B.}~\bibnamefont
  {Bauer}} \emph {et~al.},\ }\href {\doibase 10.1088/1742-5468/2011/05/P05001}
  {\bibfield  {journal} {\bibinfo  {journal} {J. Stat. Mech.}\ ,\ \bibinfo
  {pages} {P05001}} (\bibinfo {year} {2011})}\BibitemShut {NoStop}%
\end{thebibliography}%

\end{document}